\documentclass[12pt]{article}
\usepackage{amsfonts,amsmath,amssymb,mathrsfs}

\title{Quantum Hamiltonians and Stochastic Jumps}
\author{
   Detlef D\"urr\footnote{Mathematisches Institut der Universit\"{a}t
         M\"{u}nchen, Theresienstra{\ss}e 39, 80333 M\"{u}nchen, Germany.
         E-mail: duerr@mathematik.uni-muenchen.de},
   Sheldon Goldstein\footnote{Departments of Mathematics and Physics -
         Hill Center, Rutgers, The State University of New Jersey,
         110 Frelinghuysen Road, Piscataway, NJ 08854-8019, USA.
         E-mail: oldstein@math.rutgers.edu},\\
   Roderich Tumulka\footnote{Dipartimento di Fisica and INFN sezione di
         Genova,
         Via Dodecaneso 33, 16146 Genova, Italy.
         E-mail: tumulka@mathematik.uni-muenchen.de}, and
   Nino Zangh\`\i\footnote{Dipartimento di Fisica and INFN sezione di
         Genova,
         Via Dodecaneso 33, 16146 Genova, Italy. E-mail: 
zanghi@ge.infn.it}
}
\date{July 15, 2004}

\newcommand{\CCC}{\mathbb{C}} 
\newcommand{\RRR}{\mathbb{R}} 
\newcommand{\NNN}{\mathbb{N}} 
\newcommand{\ZZZ}{\mathbb{Z}} 
\newcommand{\EEE}{\mathbb{E}} 
\newcommand{\E}{e} 
\newcommand{\I}{i} 
\newcommand{\1}{\mathbf{1}} 
\newcommand{\tr}{\mathrm{tr}} 
\newcommand{\Laplace}{\Delta} 

\renewcommand{\Re}{\mathrm{Re}} 
\renewcommand{\Im}{\mathrm{Im}} 

\newcommand{\Hilbert}{\mathscr{H}}
\newcommand{\bdd}{\mathscr{B}} 
\renewcommand{\sp}[2]{\langle #1 | #2 \rangle} 
\newcommand{\scalar}[2]{\langle\!\langle #1 | #2 \rangle\!\rangle} %
\newcommand{\Fock}{\mathscr{F}} 
\newcommand{\conf}{\mathcal{Q}} 
\newcommand{\salg}{\mathcal{A}} 
\renewcommand{\div}{\,\mathrm{div}\,} 
\newcommand{\prob}{\mathrm{Prob}}
\newcommand{\measure}{\mathbb{P}} 
\newcommand{\current}{\mathbb{J}}
\newcommand{\generator}{\mathscr{L}} 
\newcommand{\pov}{{P}}
\newcommand{\profile}{\varphi}
\newcommand{\form}{{F}}
\newcommand{\domain}{\mathscr{D}}
\newcommand{\basis}{\mathscr{I}}
\newcommand{\covering}{\pi} 
\newcommand{\permutation}{\varrho} 
\newcommand{\Gommo}{\Gamma_{\!\neq}} 

\newcommand{\vx}{{\boldsymbol x}} 
\newcommand{\vy}{{\boldsymbol y}}

\newcommand{\vq}{{\boldsymbol q}}
\newcommand{\vQ}{{\boldsymbol Q}}

\newcommand{\valpha}{{\boldsymbol \alpha}}

\newcommand{\vr}{{\boldsymbol r}}

\newcommand{\inter}{{I}} 

\newcommand{\ext}{{\mathrm{ext}}} 
\newcommand{\crea}{{\mathrm{crea}}} 
\newcommand{\ann}{{\mathrm{ann}}} 

\newtheorem{theorem}{Theorem}
\newtheorem{corollary}{Corollary}

\newenvironment{proof}{\noindent 
\textit{Proof}.}{\hfill$\square$\bigskip}

\begin{document}\maketitle
\begin{abstract}
With many Hamiltonians one can naturally associate a
$|\Psi|^2$-distributed Markov process.  For nonrelativistic quantum
mechanics, this process is in fact deterministic, and is known as
Bohmian mechanics.  For the Hamiltonian of a quantum field theory, it
is typically a jump process on the configuration space of a variable
number of particles.  We define these processes for regularized
quantum field theories, thereby generalizing previous work of
John~S.~Bell \cite{BellBeables} and of ourselves \cite{crea1}.  We
introduce a formula expressing the jump rates in terms of the
interaction Hamiltonian, and establish a condition for finiteness of
the rates.

\medskip

\end{abstract}

\tableofcontents

\section{Introduction}

The central formula of this paper is
\begin{equation}\label{tranrates}
   \sigma(dq|q')= \frac{[(2/\hbar) \, \Im \, \sp{\Psi} {\pov(dq) H
   \pov(dq')| \Psi}]^+}{\sp{\Psi}{\pov(dq')| \Psi}}\,.
\end{equation}
It plays a role similar to that of Bohm's equation of motion
(\ref{Bohm}). Together, these two equations make possible a
formulation of quantum field theory (QFT) that makes no reference to
observers or measurements, while implying that observers, when making
measurements, will arrive at precisely the results that QFT is known
to predict. Special cases of formula (\ref{tranrates}) have been
utilized before \cite{BellBeables,crea1,Sudbery}. Part of what we
explain in this paper is what this formula means, how to arrive at it,
when it can be applied, and what its consequences are.  Such a
formulation of QFT takes up ideas from the seminal paper of John
S.~Bell \cite{BellBeables}, and we will often refer to theories
similar to the model suggested by Bell in \cite{BellBeables} as
``Bell-type QFTs''.  (What similar means here will be fleshed out in
the course of this paper.)

The aim of this paper is to define a canonical Bell-type model for
more or less any regularized QFT. We assume a well-defined Hamiltonian
as given; to achieve this, it is often necessary to introduce
cut-offs. We shall assume this has been done where needed.  In cases
in which one has to choose between several possible position
observables, for example because of issues related to the
Newton--Wigner operator \cite{NewtonWigner,Haag}, we shall also assume
that a choice has been made.

The primary variables of Bell-type QFTs are the positions of the
particles.  Bell suggested a dynamical law, governing the motion of
the particles, in which the Hamiltonian $H$ and the state vector
$\Psi$ determine the jump rates $\sigma$.  We point out how Bell's
rates fit naturally into a more general scheme summarized by
(\ref{tranrates}).  Since these rates are in a sense the smallest
choice possible (as explained in Section \ref{sec:mini4}), we call
them the \emph{minimal jump rates}.  By construction, they preserve
the $|\Psi|^2$ distribution.  Most of this paper concerns the
properties and mathematical foundations of minimal jump rates.  In
Bell-type QFTs, which can be regarded as extensions of Bohmian
mechanics, the stochastic jumps often correspond to the creation and
annihilation of particles. We will discuss further aspects of
Bell-type QFTs and their construction in our forthcoming work
\cite{crea2B}.

The paper is organized as follows.  In Section 2 we introduce all the
main ideas and reasonings; a superficial reading should focus on this
section.  Some examples of processes defined by minimal jump rates are
presented in Section 3.  In Section 4 we provide conditions for the
rigorous existence and finiteness of the minimal jump rates.  In
Section 5 we explain in what sense the rates \eqref{tranrates} are
minimal. Section 6 concerns further properties of processes defined by
minimal jump rates.  In Section 7 we conclude.

\section{The Jump Rate Formula}
\label{sec:making}

\subsection{Review of Bohmian Mechanics and Equivariance}

Bohmian mechanics \cite{Bohm52,DGZ,Stanford} is a non-relativistic
theory about $N$ point particles moving in 3-space, according to which
the configuration $Q=(\vQ_1,\ldots,\vQ_N)$ evolves according
to\footnote{ The masses $m_k$ of the particles have been absorbed in
the Riemann metric $g_{\mu\nu}$ on configuration space $\RRR^{3N}$,
$g_{ia,jb} = m_i \, \delta_{ij}\, \delta_{ab}$, $i,j=1\ldots N, \:
a,b=1,2,3$, and $\nabla$ is the gradient associated with $g_{\mu\nu}$,
i.e., $\nabla =(m_1^{-1}\nabla_{\vq_1}, \dots,
m_N^{-1}\nabla_{\vq_N})$.}
\begin{equation}\label{Bohm}
    \frac{dQ}{dt} = v(Q)\,,\qquad
    v=\hbar \, \Im \, \frac{\Psi^* \nabla\Psi} {\Psi^* \, \Psi}\,.
\end{equation}
$\Psi=\Psi_t(q)$ is the wave function, which
evolves according to the Schr\"odinger equation
\begin{equation}\label{Seq}
    \I\hbar\frac{\partial\Psi}{\partial t} = H \Psi\,,
\end{equation}
with
\begin{equation}\label{Hamil}
     H=  -\frac{\hbar^2}{2} \Laplace + V
\end{equation}
for spinless particles, with $\Laplace = \div\nabla$. For particles
with spin, $\Psi$ takes values in the appropriate spin space $\CCC^k$,
$V$ may be matrix valued, and numerator and denominator of
\eqref{Bohm} have to be understood as involving inner products in spin
space. The secret of the success of Bohmian mechanics in yielding the
predictions of standard quantum mechanics is the fact that the
configuration $Q_t$ is $|\Psi_t|^2$-distributed in configuration space
at all times $t$, provided that the initial configuration $Q_0$ (part
of the Cauchy data of the theory) is so distributed.  This property,
called \emph{equivariance} in \cite{DGZ}, suffices for empirical
agreement between \emph{any} quantum theory (such as a QFT) and
\emph{any} version thereof with additional (often called ``hidden'')
variables $Q$, provided the outcomes of all experiments are registered
or recorded in these variables. That is why equivariance will be our
guide for obtaining the dynamics of the particles.

The equivariance of Bohmian mechanics follows immediately from comparing
the continuity equation for a probability distribution $\rho$
associated with (\ref{Bohm}),
\begin{equation}\label{master}
   \frac{\partial \rho}{\partial t} = -\div(\rho v)\,,
\end{equation}
with the equation satisfied by $|\Psi|^2$ which follows from
(\ref{Seq}),
\begin{equation}\label{continuity1}
   \frac{\partial |\Psi|^2}{\partial t}(q,t) = \frac{2}{\hbar} \, \Im
   \, \Big[ \Psi^*(q,t)\, (H\Psi)(q,t) \Big]\,.
\end{equation}
In fact, it follows from (\ref{Hamil}) that
\begin{equation}\label{JJJ}
   \frac{2}{\hbar} \, \Im \, \Big[ \Psi^*(q,t)\, (H\Psi)(q,t) \Big]=
   -\div\Big[\hbar \, \Im \, \Psi^*(q,t) \nabla\Psi(q,t)  \Big]
\end{equation}
so, recalling (\ref{Bohm}), one obtains that
\begin{equation}\label{continuity2}
   \frac{\partial |\Psi|^2}{\partial t} =  -\div(|\Psi|^2 v)\,,
\end{equation}
and hence that if $\rho_t=|\Psi_t|^2$ at some time $t$ then
$\rho_t=|\Psi_t|^2$ for \emph{all} times.  Equivariance is an
expression of the compatibility between the Schr\"odinger evolution
for the wave function and the law, such as (\ref{Bohm}), governing the
motion of the actual configuration.  In \cite{DGZ}, in which we were
concerned only with the Bohmian dynamics \eqref{Bohm}, we spoke of the
distribution $|\Psi|^2$ as being equivariant.  Here we wish to find
processes for which we have equivariance, and we shall therefore speak
of equivariant processes and motions.

\subsection{Equivariant Markov Processes}

The study of example QFTs like that of \cite{crea1} has lead us to the
consideration of Markov processes as candidates for the equivariant
motion of the configuration $Q$ for Hamiltonians $H$ more general than
those of the form \eqref{Hamil}.

Consider a Markov process $Q_t$ on configuration space.  The
transition probabilities are characterized by the \emph{backward
generator} $L_t$, a (time-dependent) linear operator acting on
functions $f$ on configuration space:
\begin{equation}\label{backgenerator}
   L_t f(q) = \frac{d}{ds} \EEE (f(Q_{t+s})|Q_t = q)
\end{equation}
where $d/ds$ means the right derivative at $s=0$ and
$\EEE(\,\cdot\,|\,\cdot\,)$ denotes the conditional expectation.
Equivalently, the transition probabilities are characterized by the
\emph{forward generator} $\generator_t$ (or, as we shall simply say,
\emph{generator}), which is also a linear operator but acts on
(signed) measures on the configuration space.  Its defining property
is that for every process $Q_t$ with the given transition
probabilities, the distribution $\rho_t$ of $Q_t$ evolves according to
\begin{equation}\label{rhoL}
   \frac{\partial \rho_t}{\partial t} = \generator_t \rho_t\,.
\end{equation}
$\generator_t$ is the dual of $L_t$ in the sense that
\begin{equation}\label{generatorduality}
   \int f(q) \, \generator_t \rho(dq) = \int L_t f(q) \, \rho(dq)\,.
\end{equation}
We will use both $L_t$ and $\generator_t$, whichever is more
convenient.  We will encounter several examples of generators in the
subsequent sections.

We can easily extend the notion of equivariance from deterministic to
Markov processes. Given the Markov transition probabilities, we say that
\emph{the $|\Psi|^2$ distribution is equivariant} if and only if for all
times $t$ and $t'$ with $t<t'$, a configuration $Q_t$ with distribution
$|\Psi_t|^2$ evolves, according to the transition probabilities, into a
configuration $Q_{t'}$ with distribution $|\Psi_{t'}|^2$. In this case, 
we
also simply say  that the transition probabilities are
\emph{equivariant}, without explicitly mentioning $|\Psi|^2$. 
Equivariance
is equivalent to
\begin{equation}\label{genequivariance}
   \generator_t |\Psi_t|^2 = \frac{\partial |\Psi_t|^2}{\partial t}
\end{equation}
for all $t$. When \eqref{genequivariance} holds (for a fixed $t$) we
also say that $\generator_t$ is an \emph{equivariant generator} (with
respect to $\Psi_t$ and $H$). Note that this definition of
equivariance agrees with the previous meaning for deterministic
processes.

We call a Markov process $Q$ \emph{equivariant} if and only if for every
$t$ the distribution $\rho_t$ of $Q_t$ equals $|\Psi_t|^2$. For this to 
be
the case, equivariant transition probabilities are necessary but not
sufficient. (While  for a Markov process $Q$  to have equivariant
transition probabilities amounts to the property that if $\rho_t =
|\Psi_t|^2$ for one time $t$, where $\rho_t$ denotes the distribution of
$Q_t$, then $\rho_{t'} = |\Psi_{t'}|^2$ for every $t'>t$,  according to
our definition of an equivariant Markov process, in fact $\rho_t =
|\Psi_t|^2$ for all $t$.)  However, for equivariant transition
probabilities there exists a unique equivariant Markov process.

The crucial idea for our construction of an equivariant Markov process
is to note that \eqref{continuity1} is completely general, and to find
a generator $\generator_t$ such that the right hand side of
(\ref{continuity1}) can be read as the action of $\generator$ on $\rho
= |\Psi|^2$,
\begin{equation}\label{mainequ}
    \frac{2}{\hbar} \, \Im \, \Psi^* H\Psi = \generator |\Psi|^2\,.
\end{equation}
We shall implement this idea beginning in Section \ref{sec:mini1},
after a review of jump processes and some general considerations. But
first we shall illustrate the idea with the familiar case of Bohmian
mechanics.

For $H$ of the form \eqref{Hamil}, we have (\ref{JJJ}) and hence that
\begin{equation}\label{mequ}
   \frac{2}{\hbar} \, \Im \, \Psi^*H\Psi = -\div\left(\hbar \, \Im \,
   \Psi^* \nabla\Psi \right) = -\div\left( |\Psi|^2 \hbar \, \Im \,
   \frac{\Psi^* \nabla\Psi} {|\Psi|^2} \right) \,.
\end{equation}
Since the generator of the (deterministic) Markov process
corresponding to the dynamical system $dQ/dt=v(Q)$ given by a velocity
vector field $v$ is
\begin{equation}\label{dynamical}
   \generator \rho = -\div(\rho v)\,,
\end{equation}
we may recognize the last term of (\ref{mequ}) as $\generator
|\Psi|^2$ with $\generator$ the generator of the deterministic process
defined by \eqref{Bohm}. Thus, as is well known, Bohmian mechanics
arises as the natural equivariant process on configuration space
associated with $H$ and $\Psi$.

To be sure, Bohmian mechanics is not the only solution of
(\ref{mainequ}) for $H$ given by \eqref{Hamil}. Among the alternatives
are Nelson's stochastic mechanics \cite{stochmech} and other velocity
formulas \cite{Deotto}. However, Bohmian mechanics is the most natural
choice, the one most likely to be relevant to physics. (It is, in fact,
the canonical choice, in the sense of minimal process which we shall
explain in \cite[Sec.~5.2]{crea2B}.)

An important class of equivariant Markov processes are equivariant
jump processes, which we discuss in the next three sections. They
arise naturally in QFT, as we shall explain in Section~\ref{sec:2.6}.

\subsection{Equivariant Jump Processes}\label{sec:revjump}

Let $\conf$ denote the configuration space of the process,
whatever sort of space that may be (vector space, lattice, manifold,
etc.); mathematically speaking, we need that $\conf$ be a measurable
space.  A (pure) jump process is a Markov process on $\conf$ for which
the only motion that occurs is via jumps. Given that $Q_t =q$, the
probability for a jump to $q'$, i.e., into the infinitesimal volume
$dq'$ about $q'$, by time $t+dt$ is $\sigma_t(dq'|q)\, dt$, where
$\sigma$ is called the \emph{jump rate}. In this notation, $\sigma$ is
a finite measure in the first variable; $\sigma(B|q)$ is the rate (the
probability per unit time) of jumping to somewhere in the set
$B\subseteq\conf$, given that the present location is $q$. The overall
jump rate is $\sigma(\conf|q)$.

It is often the case that $\conf$ is equipped with a distinguished
measure, which we shall denote by $dq$ or $dq'$, slightly abusing
notation.  For example, if $\conf = \RRR^d$, $dq$ may be the Lebesgue
measure, or if $\conf$ is a Riemannian manifold, $dq$ may be the
Riemannian volume element. When $\sigma(\,\cdot\,|q)$ is absolutely
continuous relative to the distinguished measure, we also write
$\sigma(q'|q)\, dq'$ instead of $\sigma(dq'|q)$.  Similarly, we
sometimes use the letter $\rho$ for denoting a measure and sometimes
the density of a measure, $\rho(dq) = \rho(q)\,dq$.

A jump first occurs when a random waiting time $T$ has elapsed, after 
the
time $t_0$ at which the process was started or at which the most
recent previous jump has occurred.  For purposes of simulating or
constructing the process, the destination $q'$ can be chosen at the
time of jumping, $t_0 + T$, with probability distribution
$\sigma_{t_0+T} (\conf|q)^{-1} \, \sigma_{t_0+T} (\,\cdot\,|q)$. In
case the overall jump rate is time-independent, $T$ is exponentially
distributed with mean $\sigma(\conf|q)^{-1}$. When the
rates are time-dependent---as they will typically be in what
follows---the waiting time remains such that
\[
   \int_{t_0}^{t_0+T} \sigma_t(\conf|q) \, dt
\]
is exponentially distributed with mean 1, i.e., $T$ becomes
exponential after a suitable (time-dependent) rescaling of time. For
more details about jump processes, see \cite{Breiman}.

The generator of a pure jump process can be expressed in terms of the
rates:
\begin{equation}\label{continuity3}
   \generator_\sigma \rho(dq) = \int\limits_{q'\in\conf}  \Big(
   \sigma(dq|q') \rho(dq') - \sigma(dq'|q) \rho(dq) \Big)\,,
\end{equation}
a ``balance'' or ``master'' equation expressing $\partial
\rho/\partial t$ as the gain due to jumps to $dq$ minus the loss due
to jumps away from $q$.

We shall say that jump rates $\sigma$ are \emph{equivariant} if
$\generator_\sigma$ is an equivariant generator.  It is one of our goals
in
this paper to describe  a general scheme for obtaining equivariant jump
rates. In Sections \ref{sec:mini1} and \ref{sec:mini2} we will explain 
how
this leads us to formula \eqref{tranrates}.

\subsection{Integral Operators Correspond to Jump Processes}
\label{sec:mini1}

What characterizes jump processes versus continuous processes is that
some amount of probability that vanishes at $q\in\conf$ can reappear
in an entirely different region of configuration space, say at
$q'\in\conf$. This is manifest in the equation for $\partial
\rho/\partial t$, (\ref{continuity3}): the first term in the integrand
is the probability increase due to arriving jumps, the second the
decrease due to departing jumps, and the integration over $q'$
reflects that $q'$ can be anywhere in $\conf$. This suggests that
Hamiltonians for which the expression \eqref{continuity1} for
$\partial |\Psi|^2/\partial t$ is naturally an integral over $dq'$
correspond to pure jump processes. So when is the left hand side of
(\ref{mainequ}) an integral over $dq'$?  When $H$ is an integral
operator, i.e., when $\sp{q}{H|q'}$ is not merely a formal symbol, but
represents an integral kernel that exists as a function or a measure and
satisfies
\begin{equation}
   (H\Psi)(q) = \int dq'\,\sp{q}{H|q'}\, \Psi(q')\,.
\end{equation}
(For the time being, think of $\conf$ as $\RRR^d$ and of wave
functions as complex valued.) In this case, we should choose the jump
rates in such a way that, when $\rho = |\Psi|^2$,
\begin{equation}\label{la1}
   \sigma(q|q') \,\rho(q') - \sigma(q'|q) \,\rho(q) = \frac{2}{\hbar}
   \, \Im \, \Psi^*(q)\, \sp{q}{H|q'} \, \Psi(q') \,,
\end{equation}
and this suggests, since jump rates must be nonnegative (and the right
hand side of \eqref{la1} is anti-symmetric), that
\[
   \sigma(q|q') \,\rho(q') = \Big[ \frac{2}{\hbar} \, \Im \,
   \Psi^*(q)\, \sp{q}{H|q'} \, \Psi(q') \Big]^+
\]
(where $x^+$ denotes the positive part of $x\in\RRR$, that is, $x^+$ is
equal to $x $ for $x>0$ and is zero otherwise), or
\begin{equation}\label{mini1}
   \sigma(q|q') = \frac{ \big[ (2/\hbar) \, \Im \, \Psi^*(q) \, \sp{q}
   {H|q'} \, \Psi(q') \big]^+}{\Psi^*(q')\, \Psi(q')} .
\end{equation}
These rates are an instance of what we call the \emph{minimal jump
rates} associated with $H$ (and $\Psi$).  They are also an instance of
formula (\ref{tranrates}), as will become clear in the following
section.  The name comes from the fact that they are actually the
minimal possible values given (\ref{la1}), as is expressed by the
inequality \eqref{minimality} and will be explained in detail in
Section \ref{sec:mini4}.  Minimality entails that at any time $t$, one
of the transitions $q_1 \to q_2$ or $q_2 \to q_1$ is forbidden. We
will call the process defined by the minimal jump rates the
\emph{minimal jump process} (associated with $H$).

In contrast to jump processes, continuous motion, as in Bohmian
mechanics, corresponds to such Hamiltonians that the formal matrix
elements $\sp{q}{H|q'}$ are nonzero only infinitesimally close to the
diagonal, and in particular to differential operators like the
Schr\"odinger Hamiltonian (\ref{Hamil}), which has matrix elements of
the type $\delta''(q-q') + V(q) \,\delta(q-q')$.

The minimal jump rates as given by (\ref{mini1}) have some nice
features.  The possible jumps for this process correspond to the
nonvanishing matrix elements of $H$ (though, depending on the state
$\Psi$, even some of the jump rates corresponding to nonvanishing
matrix elements of $H$ might happen to vanish).  Moreover, in their
dependence on the state $\Psi$, the jump rates $\sigma$ depend only
``locally'' upon $\Psi$: the jump rate for a given jump $q'\to q$
depends only on the values $\Psi(q')$ and $\Psi(q)$ corresponding to
the configurations linked by that jump.  Discretizing $\RRR^3$ to a
lattice $\varepsilon \ZZZ^3$, one can obtain Bohmian mechanics as a
limit $\varepsilon\to 0$ of minimal jump processes
\cite{Sudbery,Vink}, whereas greater-than-minimal jump rates lead to
Nelson's stochastic mechanics \cite{stochmech} and similar diffusions;
see \cite{Vink,Guerra}. If the Schr\"odinger operator \eqref{Hamil} is
approximated in other ways by operators corresponding to jump
processes, e.g., by $H_\varepsilon = \E^{-\varepsilon H} H
\E^{-\varepsilon H}$, the minimal jump processes presumably also
converge to Bohmian mechanics.

We have reason to believe that there are lots of self-adjoint
operators which do not correspond to any stochastic process that can
be regarded as defined, in any reasonable sense, by
\eqref{mini1}.\footnote{Consider, for example, $H = p \cos p$ where
$p$ is the one-dimensional momentum operator $-\I\hbar
\partial/\partial q$. Its formal kernel $\sp{q}{H|q'}$ is the
distribution $-\frac{\I}{2} \delta'(q-q'-1) - \frac{\I}{2}
\delta'(q-q'+1)$, for which \eqref{mini1} would not have a meaning.
{}From a sequence of smooth functions converging to this distribution,
one can obtain a sequence of jump processes with rates \eqref{mini1}:
the jumps occur very frequently, and are by amounts of approximately
$\pm 1$.  A limiting process, however, does not exist.}  But such
operators seem never to occur in QFT. (The Klein--Gordon operator
$\sqrt{m^2 c^4 - \hbar^2 c^2 \Laplace}$ does seem to have a process,
but it requires a more detailed discussion which will be provided in a
forthcoming work \cite{klein2}.)

\subsection{Minimal Jump Rates}
\label{sec:mini2}

The reasoning of the previous section applies to a far more general
setting than just considered: to arbitrary configuration spaces
$\conf$ and ``generalized observables''---POVMs---defining, for our
purposes, what the ``position representation'' is. We now present this
more general reasoning, which leads to formula (\ref{tranrates}).

The process we construct relies on the following ingredients from QFT:
\begin{enumerate}
\item A Hilbert space $\Hilbert$ with scalar product $\sp{\Psi}
   {\Phi}$.

\item A unitary one-parameter group $U_t$ in $\Hilbert$ with
   Hamiltonian $H$,
   \[
     U_t = \E^{-\frac{\I}{\hbar}tH}\,,
   \]
   so that in the Schr\"odinger picture the state $\Psi$ evolves
   according to
   \begin{equation}
     \I\hbar\frac{d\Psi_t}{dt} = H\Psi_t\,.
   \end{equation}
   $U_t$ could be part of a representation of the Poincar\'e group.

\item A positive-operator-valued measure (POVM) $\pov(dq)$ on $\conf$
   acting on $\Hilbert$, so that the probability that the system in the
   state $\Psi$ is localized in $dq$ at time $t$ is
   \begin{equation} \label{mis}
     \measure_t(dq)= \sp{\Psi_t}{\pov(dq)| \Psi_t} \,.
   \end{equation}
\end{enumerate}

Mathematically, a POVM $\pov$ on $\conf$ is a countably additive set
function (``measure''), defined on measurable subsets of $\conf$, with
values in the positive (bounded self-adjoint) operators on (a Hilbert
space) $\Hilbert$, such that $\pov(\conf)$ is the identity
operator.\footnote{The countable additivity is to be understood as in
the sense of the weak operator topology. This in fact implies that
countable additivity also holds in the strong topology.}  Physically,
for our purposes, $\pov(\,\cdot\,)$ represents the (generalized)
position observable, with values in $\conf$.  The notion of POVM
generalizes the more familiar situation of observables given by a set
of commuting self-adjoint operators, corresponding, by means of the
spectral theorem, to a projection-valued measure (PVM): the case where
the positive operators are projection operators.  A typical example is
the single Dirac particle: the position operators on
$L^2(\RRR^3,\CCC^4)$ induce there a natural PVM $\pov_0(\,\cdot\,)$:
for any Borel set $B\subseteq \RRR^3$, $\pov_0(B)$ is the projection
to the subspace of functions that vanish outside $B$, or,
equivalently, $\pov_0(B)\Psi(q) = \1_B(q) \, \Psi(q)$ with $\1_B$ the
indicator function of the set $B$.  Thus, $\sp{\Psi} {\pov_0 (dq)|
\Psi} = |\Psi(q)|^2 dq$.  When one considers as Hilbert space
$\Hilbert$ only the subspace of positive energy states, however, the
localization probability is given by $\pov(\,\cdot\,) = P_+
\pov_0(\,\cdot\,) I$ with $P_+:L^2(\RRR^3,\CCC^4) \to \Hilbert$ the
projection and $I:\Hilbert \to L^2(\RRR^3,\CCC^4)$ the inclusion
mapping. Since $P_+$ does not commute with most of the operators
$\pov_0(B)$, $\pov (\,\cdot\,)$ is no longer a PVM but a genuine
POVM\footnote{This situation is indeed more general than it may seem.
By a theorem of Naimark \cite[p.~142]{Davies}, every POVM $\pov
(\,\cdot\,)$ acting on $\Hilbert$ is of the form $\pov(\,\cdot\,) =
P_+ \pov_0 (\,\cdot\,) P_+$ where $\pov_0$ is a PVM on a larger
Hilbert space, and $P_+$ the projection to
$\Hilbert$. \label{ft:Naimark}} and consequently does not correspond
to any position operator---although it remains true (for $\Psi$ in the
positive energy subspace) that $\sp{\Psi}{\pov(dq)| \Psi} =
|\Psi(q)|^2 dq$.  That is why in QFT, the position observable is
indeed more often a POVM than a PVM. POVMs are also relevant to
photons \cite{ali,kraus}.  In one approach, the photon wave function
$\Psi: \RRR^3 \to \CCC^3$ is subject to the constraint condition
$\nabla \cdot \Psi = \partial_1 \Psi_1 + \partial_2 \Psi_2 +
\partial_3 \Psi_3 =0$. Thus, the physical Hilbert space $\Hilbert$ is
the (closure of the) subspace of $L^2(\RRR^3,\CCC^3)$ defined by this
constraint, and the natural PVM on $L^2(\RRR^3,\CCC^3)$ gives rise, by
projection, to a POVM on $\Hilbert$.  So much for POVMs. Let us get
back to the construction of a jump process.

The goal is to specify equivariant jump rates $\sigma = \sigma^{\Psi, H,
\pov}$, i.e., such rates that
\begin{equation}\label{equirates}
   \generator_\sigma \measure = \frac{d\measure}{dt} \,.
\end{equation}
To this end, one may take the following steps:

\begin{enumerate}
\item Note that
   \begin{equation}\label{dPdt}
     \frac{d\measure_t(dq)}{dt} = \frac{2}{\hbar} \, \Im \,
     \sp{\Psi_t}{\pov(dq) H| \Psi_t}\,.
   \end{equation}
\item Insert the resolution of the identity $I = 
\int\limits_{q'\in\conf}
   \pov(dq')$ and obtain
   \begin{equation}\label{dPdtJ}
     \frac{d\measure_t(dq)}{dt} =\int\limits_{q'\in\conf}
     \current_t(dq,dq') \,,
   \end{equation}
   where
   \begin{equation}\label{Jdef}
     \current_t(dq,dq') = \frac{2}{\hbar} \,
     \Im \, \sp{\Psi_t}{\pov(dq)H \pov(dq')| \Psi_t} \,.
   \end{equation}
\item Observe that $\current$ is anti-symmetric, $\current(dq',dq) = -
   \current(dq,dq')$.  Thus, since $x = x^+ - (-x)^+$,
   \[
     \current(dq,dq') = \left[(2/\hbar) \, \Im \, \sp{\Psi} {\pov(dq) H
     \pov(dq') |\Psi}\right]^+ - \left[(2/\hbar)\, \Im \, \sp{\Psi}
     {\pov(dq') H \pov(dq) |\Psi}\right]^+ .
   \]
\item Multiply and divide both terms by $\measure(\,\cdot\,)$,
   obtaining that
   \begin{eqnarray*}
     \int\limits_{q'\in\conf} \current(dq,dq') = \int\limits_{q'\in\conf}
     \bigg( \hspace{-3ex} &&
     \frac{[(2/\hbar) \, \Im \, \sp{\Psi} {\pov(dq) H \pov(dq')| 
\Psi}]^+}
     {\sp{\Psi}{\pov(dq')| \Psi}} \measure(dq') -
   \\-&&
     \frac{[(2/\hbar) \, \Im \, \sp{\Psi} {\pov(dq') H \pov(dq)| \Psi}
     ]^+} {\sp{\Psi} {\pov(dq)| \Psi}} \measure(dq) \bigg) \,.
\end{eqnarray*}
\item By comparison with \eqref{continuity3}, recognize the right hand
   side of the above equation as $\generator_\sigma \measure$, with
   $\generator_\sigma$ the generator of a Markov jump process with jump
   rates \eqref{tranrates}, which we call the \emph{minimal jump
   rates}. We repeat the formula for convenience:
   \[
     \sigma(dq|q')= \frac{[(2/\hbar) \, \Im \, \sp{\Psi} {\pov(dq) H
     \pov(dq')| \Psi}]^+}{\sp{\Psi}{\pov(dq')| \Psi}}\,.
   \]
\end{enumerate}
Mathematically, the right hand side of this formula as a function of
$q'$ must be understood as a density (Radon--Nikod{\'y}m derivative)
of one measure relative to another. The plus symbol denotes the
positive part of a signed measure; it can also be understood as
applying the plus function, $x^+ = \max (x,0)$, to the density, if it
exists, of the numerator.

To sum up, we have argued that with $H$ and $\Psi$ is naturally
associated a Markov jump process $Q_t$ whose marginal distributions
coincide at all times by construction with the quantum probability
measure, $\rho_t(\,\cdot\,) = \measure_t(\,\cdot\,)$, so that $Q_t$ is
an equivariant Markov process.

In Section \ref{sec:maths}, we establish precise conditions on
$H,\pov$, and $\Psi$ under which the jump rates \eqref{tranrates} are
well-defined and finite $\measure$-almost everywhere, and prove that
in this case the rates are equivariant, as suggested by the steps 1-5
above. It is perhaps worth remarking at this point that any $H$ can be
approximated by Hamiltonians $H_n$ (namely Hilbert--Schmidt operators)
for which the rates \eqref{tranrates} are always (for all $\Psi$)
well-defined and equivariant, as we shall prove in Section
\ref{sec:HS}.

\subsection{Bell-Type QFT}
\label{sec:2.6}\label{sec:BellQFT}

A Bell-type QFT is about particles moving in physical 3-space; their
number and positions are represented by a point $Q_t$ in configuration
space $\conf$, with $\conf$ defined as follows. Let $\Gamma \RRR^3$
denote the configuration space of a variable (but finite) number of
identical particles in $\RRR^3$, i.e., the union of $(\RRR^3)^n$
modulo permutations,
\begin{equation}
  \Gamma \RRR^3 = \bigcup_{n=0}^\infty (\RRR^3)^n /S_n \,.
\end{equation}
$\conf$ is the Cartesian product of several copies of $\Gamma \RRR^3$,
one for each species of particles. For a discussion of the space
$\Gamma \RRR^3$, and indeed of $\Gamma S$ for any other
measurable space $S$ playing the role of physical space, see
\cite[Sec.~2.8]{crea2B}.

A related space, for which we write $\Gommo \RRR^3$, is the space of
all finite subsets of $\RRR^3$; it is contained in $\Gamma \RRR^3$,
after obvious identifications. In fact, $\Gommo \RRR^3 = \Gamma \RRR^3
\setminus \Delta$, where $\Delta$ is the set of coincidence
configurations, i.e., those having two or more particles at the same
position.  $\Gommo \RRR^3$ is the union of the spaces
${\conf}^{(n)}_{\neq}$ for $n=0,1,2, \ldots$, where
${\conf}^{(n)}_{\neq}$ is the space of subsets of $\RRR^3$ with $n$
elements, a manifold of dimension $3n$ (see \cite{identical} for a
discussion of Bohmian mechanics on this manifold). The set $\Delta$ of
coincidence configurations has codimension $3$ and thus can usually be
ignored.  We can thus replace $\Gamma \RRR^3$ by the somewhat simpler
space $\Gommo \RRR^3$.

$Q_t$ follows a Markov process in $\conf$, which is governed by a
state vector $\Psi$ in a suitable Hilbert space $\Hilbert$.
$\Hilbert$ is related to $\conf$ by means of a PVM or POVM $\pov$.

The Hamiltonian of a QFT usually comes as a sum, such as
\begin{equation}\label{Hsum}
  H = H_0 + H_\inter
\end{equation}
with $H_0$ the free Hamiltonian and $H_\inter$ the interaction
Hamiltonian. If several particle species are involved, $H_0$ is itself
a sum containing one free Hamiltonian for each species. The left hand
side of (\ref{mainequ}), which should govern our choice of the
generator, is then also a sum,
\begin{equation}\label{Hsumgen}
  \frac{2}{\hbar} \, \Im \, \Psi^* H_0 \Psi + \frac{2}{\hbar} \, \Im
  \, \Psi^* H_\inter \Psi = \generator |\Psi|^2\,.
\end{equation}
This opens the possibility of finding a generator $\generator$ by
setting $\generator = \generator_0 + \generator_\inter$, provided we
have generators $\generator_0$ and $\generator_\inter$
corresponding to $H_0$ and $H_\inter$ in the sense that
\begin{subequations}
\begin{align}
  \frac{2}{\hbar} \, \Im \, \Psi^* H_0 \Psi
  &= \generator_0 |\Psi|^2 \\
  \frac{2}{\hbar} \, \Im \, \Psi^* H_\inter \Psi
  &= \generator_\inter |\Psi|^2\,.
\end{align}
\end{subequations}
This feature of (\ref{mainequ}) we call \emph{process additivity}; it
is based on the fact that the left hand side of (\ref{mainequ}) is
linear in $H$.

In a Bell-type QFT, the generator $\generator$ is of the form
$\generator = \generator_0 + \generator_I$, where $\generator_0$ is
usually the generator of a deterministic process, usually defined by
the Bohmian or Bohm--Dirac law of motion, see below, and
$\generator_I$ is the generator of a pure jump process, which is our
main focus in this paper.  The process generated by $\generator$ is
then given by deterministic motion determined by $\generator_0$,
randomly interrupted by jumps at a rate determined by $\generator_I$.

We thus need to define two equivariant processes, one (the ``free
process'') associated with $H_0$ and the other (the ``interaction
process'') with $H_I$. The interaction process is the pure jump
process with rates given by \eqref{tranrates} with $H_I$ in place
of $H$.  We now give a description of the free process for the two
most relevant free Hamiltonians: the second-quantized Schr\"odinger
operator and the second-quantized Dirac operator. We give a more
general and more detailed discussion of free processes in
\cite{crea2B}; there we provide a formula, roughly analogous to
\eqref{tranrates}, for $\generator_0$ in terms of $H_0$, and an
algorithm for obtaining the free process from a one-particle process
that is roughly analogous to the ``second quantization'' procedure for
obtaining $H_0$ from a one-particle Hamiltonian.

The free process associated with a second-quantized Schr\"odinger
operator arises from Bohmian mechanics.  Fock space $\Hilbert = \Fock$
is a direct sum
\begin{equation}\label{fockspace}
  \Fock= \bigoplus_{n=0}^{\infty} \Fock^{(n)} ,
\end{equation}
where $\Fock^{(n)}$ is the $n$-particle Hilbert space.  $\Fock^{(n)}$
is the subspace of symmetric (for bosons) or anti-symmetric (for
fermions) functions in $L^2 (\RRR^{3n}, (\CCC^{2s+1})^{\otimes n})$
for spin-$s$ particles. Thus, $\Psi \in \Fock$ can be decomposed into
a sequence $\Psi = \left( \Psi^{(0)}, \Psi^{(1)}, \ldots, \Psi^{(n)},
\ldots \right)$, the $n$-th member $\Psi^{(n)}$ being an $n$-particle
wave function, the wave function representing the $n$-particle sector
of the quantum state vector.  The obvious way to obtain a process on
$\conf = \Gamma \RRR^3$ is to let the configuration $Q(t)$, containing
$N = \#Q(t)$ particles, move according to the $N$-particle version of
Bohm's law (\ref{Bohm}), guided by $\Psi^{(N)}$.\footnote{As defined,
configurations are unordered, whereas we have written Bohm's law
\eqref{Bohm} for ordered configurations.  Thanks to the
(anti\nobreakdash-)symmetry of the wave function, however, all
orderings will lead to the same particle motion. For more about such
considerations, see our forthcoming work \cite{identical}.}  This is
indeed an equivariant process since $H_0$ has a block diagonal form
with respect to the decomposition (\ref{fockspace}),
\[
  H_0 = \bigoplus_{n=0}^\infty H_0^{(n)}\,,
\]
and $H_0^{(n)}$ is just a Schr\"odinger operator for $n$
noninteracting particles, for which, as we already know, Bohmian
mechanics is equivariant.  We used a very similar process in
\cite{crea1} (the only difference being that particles were numbered
in \cite{crea1}).

Similarly, if $H_0$ is the second quantized Dirac operator, we let a
configuration $Q$ with $N$ particles move according to the usual
$N$-particle Bohm--Dirac law \cite[p.~274]{BH}
\begin{equation}\label{BohmDirac}
  \frac{dQ}{dt} = c\frac{\Psi^*(Q) \, \alpha_{N} \, \Psi(Q)}
  {\Psi^*(Q) \, \Psi(Q)}
\end{equation}
where $c$ denotes the speed of light and $\alpha_{N} = (\valpha^{(1)},
\ldots, \valpha^{(N)})$ with $\valpha^{(k)}$ acting on the spin index
of the $k$-th particle.

This completes the construction of the Bell-type QFT. An explicit
example of a Bell-type process for a simple QFT is described in
\cite{crea1}, which we take up again in Section~\ref{sec:crea1} below
to point out how its jump rates fit into the scheme
\eqref{tranrates}. Another such example, concerning electron--positron
pair creation in an external electromagnetic field, is described in
\cite[Sec.~3.3.]{crea2B}.

%

\section{Examples}
\label{sec:example}

In this section, we present various special cases of the jump rate
formula \eqref{tranrates} and examples of its application. We also
point out how the jump rates of the models in \cite{crea1} and
\cite{BellBeables} are contained in \eqref{tranrates}.

\subsection{A First Example}\label{sec:ex1}

To begin with, we consider $\conf = \RRR^d$, $\Hilbert =
L^2(\RRR^d,\CCC)$, and $\pov$ the natural PVM, which may be written
$\pov(dq) = |q\rangle \langle q | \, dq$. Then, $\measure(dq) =
\sp{\Psi}{\pov(dq) | \Psi} = |\Psi(q)|^2 dq$, and the jump rate
formula \eqref{tranrates} reads
\begin{subequations}
\begin{align}
   \sigma(q|q') &= \frac{[(2/\hbar) \, \Im \, \Psi^*(q) \, \sp{q}{H|q'}
   \, \Psi(q')]^+}{\Psi^*(q') \, \Psi(q')} \label{Rdrates}\\
   &= \Big[ \frac{2}{\hbar} \, \Im \, \frac{ \Psi^*(q) \,
   \sp{q}{H|q'}}{\Psi^*(q')} \Big]^+ .
\end{align}
\end{subequations}
Note that \eqref{Rdrates} is the same expression as \eqref{mini1}.  As
a simple example of an operator $H_\inter$ with a kernel, consider a
convolution operator, $H_\inter = V \star$, where $V$ may be
complex-valued and  $V(-q) = V^*(q)$,
\[
   (H_\inter\Psi)(q)=\int V(q-q') \, \Psi(q') \, dq'\,.
\]
The kernel is $\sp{q}{H_\inter|q'} = V(q-q')$. Together with $H_0 =
-\frac{\hbar^2}{2} \Laplace$, we obtain a baby example of a
Hamiltonian $H = H_0 + H_\inter$ that goes beyond the form
(\ref{Hamil}) of Schr\"odinger operators, in particular in that it is
no longer local in configuration space. Recall that $H_0$ is
associated with the Bohmian motion (\ref{Bohm}). Combining the two
generators on the basis of process additivity, we obtain a process
that is piecewise deterministic, with jump rates (\ref{mini1}) and
Bohmian trajectories between successive jumps.

\subsection{Wave Functions with Spin}\label{sec:exspin}

Let us next become a bit more general and consider wave functions with
spin, i.e., with values in $\CCC^k$. We have $\conf = \RRR^d, \Hilbert
= L^2(\RRR^d,\CCC^k)$ and $\pov$ the natural PVM, which may be written
$\pov(dq) = \sum_{i=1}^k |q,i\rangle \langle q,i| \, dq$, where $i$
indexes the standard basis of $\CCC^k$. Another way of viewing $\pov$
is to understand $\Hilbert$ as the tensor product $L^2(\RRR^d,\CCC)
\otimes \CCC^k$, and $\pov(dq) = \pov_0(dq) \otimes I_{\CCC^k}$ with
$\pov_0$ the natural PVM on $L^2(\RRR^d,\CCC)$ and $I_{\CCC^k}$ the
identity operator on $\CCC^k$. Using the notation
$\scalar{\Phi(q)}{\Psi(q)}$ for the scalar product in $\CCC^k$, we can
write $\measure(dq) = \sp{\Psi}{\pov(dq)| \Psi} =
\scalar{\Psi(q)}{\Psi(q)} \, dq$, and the jump rate formula
\eqref{tranrates} reads
\begin{equation}\label{multicomponent}
   \sigma(q|q') = \frac{[(2/\hbar) \, \Im \,
   \scalar{\Psi(q)}{K(q,q')|\Psi(q')}]^+}{\scalar{\Psi(q')}{\Psi(q')}}
\end{equation}
with $K(q,q')$, the kernel of $H$, a $k \times k$ matrix. If we write
$\Phi^*(q) \, \Psi(q)$ for $\scalar{\Phi(q)}{\Psi(q)}$, as we did in
\eqref{Bohm} and \eqref{BohmDirac}, and $\sp{q}{H|q'}$ for $K(q,q')$,
\eqref{multicomponent} reads
\[
   \sigma(q|q') = \frac{[(2/\hbar) \, \Im \, \Psi^*(q) \, \sp{q}{H|q'}
   \, \Psi(q')]^+}{\Psi^*(q') \, \Psi(q')}\,,
\]
which is \eqref{mini1} again, interpreted in a different way.

\subsection{Vector Bundles}\label{sec:exbundle}

Next consider, instead of the fixed value space $\CCC^k$, a vector
bundle $E$ over a Riemannian manifold $\conf$, and cross-sections of
$E$ as wave functions.  In order to have a scalar product of wave
functions, we need that every bundle fiber $E_q$ be equipped with a
Hermitian inner product $\scalar{\,\cdot\,}{\,\cdot\,}_q$.  We
consider $\Hilbert = L^2(E)$ (the space of square-integrable
cross-sections) and $\pov$ the natural PVM. For any $q$ and $q'$,
$K(q,q')$ then has to be a $\CCC$-linear mapping $E_{q'} \to E_q$, so
that the kernel of $H$ is a cross-section of the bundle
$\bigcup_{q,q'} E_q \otimes E^*_{q'}$ over $\conf \times \conf$.
\eqref{tranrates} then reads
\begin{equation}\label{bundlerates}
   \sigma(q|q') = \frac{[(2/\hbar) \, \Im \, \scalar{\Psi(q)} {K(q,q')
   \, \Psi(q')}_q]^+} {\scalar{\Psi(q')} {\Psi(q')}_{q'}}\,.
\end{equation}
In the following we will use the notation $\Phi^*(q) \, \Psi(q)$ for
$\scalar{\Phi(q)}{\Psi(q)}_q$ and $\sp{q}{H|q'}$ for $K(q,q')$, so
that
\[
   \sigma(q|q') = \frac{[(2/\hbar) \, \Im \, \Psi^*(q) \, \sp{q}{H|q'}
   \, \Psi(q')]^+}{\Psi^*(q') \, \Psi(q')}\,,
\]
which looks like \eqref{mini1} again.

\subsection{Kernels of the Measure Type}\label{sec:ex3}

The kernel $\sp{q}{H|q'}$ can be less regular than a function.  Since 
the
numerator of \eqref{tranrates} is a measure in $q$ and $q'$, the formula
still makes sense (for $P$ the natural PVM)  when the kernel
$\sp{q}{H|q'}$ is a complex measure in $q$ and $q'$. The mathematical
details will be discussed in Section \ref{sec:jumps}. For instance, the
kernel can have singularities like a Dirac $\delta$, but it cannot have
singularities worse than $\delta$, such as derivatives of $\delta$ (as
would arise from an operator whose position representation is a
differential operator).  It can happen that the kernel is not a function
but a measure even for a very well-behaved (even bounded) operator. For
example,  this is the case for $H$  a multiplication operator (i.e., a
function $V(\hat{q})$ of the position operator), $\sp{q}{H|q'} = V(q) \,
\delta(q-q')$. Note, though, that multiplication operators correspond to
zero jump rates.

A nontrivial example of an operator with $\delta$ singularities in the
kernel is $H = 1-\cos (p/p_0)$ where $p= -\I\hbar \partial/\partial q$
is the momentum operator in one dimension, $\Hilbert =
L^2(\RRR,\CCC)$, and $p_0$ is  a constant. The dispersion relation $E =
1-\cos (p/p_0)$ begins at $p=0$ like $\frac{1}{2}(p/p_0)^2$ but
deviates from the parabola for large $p$. In the position
representation, $H$ is the convolution with ($(2\pi)^{-1/2}$ times)
the inverse Fourier transform of the function $1-\cos (\hbar k/p_0)$,
and thus $\sp{q}{H|q'} = \delta(q-q') - \frac{1}{2} \, \delta(q-q'+
\frac{\hbar}{p_0}) -\frac{1}{2} \, \delta(q-q'- \frac{\hbar}{p_0})$. In
this case, \eqref{tranrates} leads to
\begin{equation}\label{ex3rates}
   \sigma(q|q') = \frac{[(-1/\hbar) \, \Im \, \Psi^*(q) \,
   \Psi(q')]^+}{\Psi^*(q') \, \Psi(q')} \, \Big(
   \delta(q-q'+\tfrac{\hbar}{p_0}) + \delta(q-q'- \tfrac{\hbar}{p_0})
   \Big).
\end{equation}
(Note that nonnegative factors can be drawn out of the plus function.)
This formula may be viewed as contained in \eqref{mini1} as well, in a
formal sense.  As a consequence of \eqref{ex3rates}, only jumps by an
amount of $\pm \frac{\hbar}{p_0}$ can occur in this case.

\subsection{Infinite Rates}\label{sec:inftyrates}

There also exist Markov processes that perform \emph{infinitely many
jumps} in every finite time interval (e.g., Glauber dynamics for
infinitely many spins). These processes, which we do not count among
the jump processes, may appear pathological, and we will not
investigate them in this paper, but we note that some Hamiltonians may
correspond to such processes. They could arise from jump rates
$\sigma(\,\cdot\,|q')$ given by \eqref{tranrates} that form not a
finite but merely a $\sigma$-finite measure, so that $\sigma(\conf|q')
= \infty$. Here is an (artificial) example of $\sigma$-finite (but not
finite)  rates,
arising from an operator $H$ that is even bounded.

Let $\conf = \RRR$, $\Hilbert = L^2(\RRR)$ with $\pov(\,\cdot\,)$ the
position PVM, and let  $H$, in Fourier representation, be
multiplication by $f(k) = \sqrt{\pi/2} \, \mathrm{sign}(k)$. $H$ is
bounded since $f$ is. $f$ is the Fourier transform of $\I/x$,
understood as the distribution defined by the principal value
integral.  As a consequence, $H$ has, in position representation, the
kernel $\sp{q}{H|q'} = \I/(q-q')$. {}From \eqref{mini1} we obtain the
jump rates
\begin{equation}\label{inftyrates}
   \sigma(q|q') = \frac{2}{\hbar} \frac{1}{\Psi^*(q') \, \Psi(q')}
   \Big[ \frac{\Re \, \Psi^*(q) \, \Psi(q')}{q-q'} \Big]^+ \,,
\end{equation}
which entails that $\sigma(\RRR|q') = \int \sigma(q|q') \, dq =
\infty$ at least whenever $\Psi$ is continuous (and nonvanishing) at
$q'$. Nonetheless, since the rate for jumping anywhere outside the
interval $[q' - \varepsilon, q' + \varepsilon]$ is finite for every
$\varepsilon >0$ and since $\int_{q'- \varepsilon}^{q' + \varepsilon}
|q-q'|\sigma(q|q') \, dq < \infty$, a process with these rates should
exist: among the jumps that the process would have to make per unit
time, the large ones would be few and the frequent ones would be
tiny---too tiny to significantly contribute.

\subsection{Discrete Configuration Space}
\label{sec:discrete}

Now consider a discrete configuration space $\conf$. Mathematically,
this means $\conf$ is a countable set. In this case, measures are
determined by their values on singletons $\{q\}$, and we can specify
all jump rates by specifying the rate $\sigma(q|q')$ for each
transition $q' \to q$. \eqref{tranrates} then reads
\begin{equation}\label{disrates2}
   \sigma(q|q') = \frac{\big[ (2/\hbar) \, \Im \, \sp{\Psi} {\pov\{q\}
   H \pov\{q'\}| \Psi} \big]^+} {\sp{\Psi}{\pov\{q'\}| \Psi}} \,.
\end{equation}

  We begin with the particularly simple case that there is an
orthonormal basis of $\Hilbert$ labeled by $\conf$, $\{ |q\rangle :
q\in\conf \}$, and $\pov$ is the PVM corresponding to this basis,
$\pov\{q\} = |q\rangle \langle q|$. In this case, the notation
$\sp{q}{H|q'}$ and the name ``matrix element'' can be taken
literally. The rates \eqref{tranrates} then simplify to
\begin{subequations}
\begin{align}
   \sigma(q|q') &= \frac{[(2/\hbar) \, \Im \, \sp{\Psi}{q} \sp{q}{H|q'}
   \sp{q'}{\Psi}]^+}{\sp{\Psi}{q'} \sp{q'}{\Psi}} \label{disrates1}\\
   &= \Big[ \frac{2}{\hbar} \, \Im \, \frac{ \sp{\Psi}{q}
   \sp{q}{H|q'}}{\sp{\Psi}{q'}} \Big]^+ .
\end{align}
\end{subequations}
Note that \eqref{disrates1} is the obvious discrete analogue of
\eqref{mini1}; in fact, one can regard \eqref{mini1} as another way
of writing \eqref{disrates1} in this case.

Consider now the more general case that a basis of Hilbert space is
indexed by two ``quantum numbers,'' the configuration $q$ and another
index $i$.  Then the POVM is given by the PVM $\pov\{q\} = \sum_i
|q,i\rangle \langle q,i |$, the projection onto the subspace
associated with $q$ (whose dimension might depend on $q$); such a PVM
may be called ``degenerate.'' We have $\measure(q) =
\sp{\Psi}{\pov\{q\} |\Psi} = \sum\limits_i \sp{\Psi}{q,i}
\sp{q,i}{\Psi}$, and \eqref{tranrates} becomes
\begin{equation}\label{degenrates}
   \sigma(q|q') = \frac{\Big[\frac{2}{\hbar} \, \Im \, \sum\limits_{i,i'}
   \sp{\Psi}{q,i} \sp{q,i}{H|q',i'} \sp{q',i'}{\Psi} \Big]^+}
   {\sum\limits_{i'} \sp{\Psi}{q',i'} \sp{q',i'}{\Psi}} \,.
\end{equation}
We may also write \eqref{degenrates} as \eqref{disrates1},
understanding $\sp{\Psi}{q}$ and $\sp{q'}{\Psi}$ as multi-component,
$\sp{q}{H|q'}$ as a matrix, and products as inner products.  In case
that the dimension of the subspace associated with $q$ is always $k$,
independent of $q$, \eqref{degenrates} is a discrete analogue of the
rate formula \eqref{multicomponent} for spinor-valued wave functions.

Apart from serving as mathematical examples, discrete configuration
spaces are relevant for several reasons: First, they provide
particularly simple cases of jump processes with minimal rates that
are easy to study. Second, any \emph{numerical} computation is discrete 
by
nature. Third, one may consider approximating or replacing the
$\RRR^3$ that is supposed to model physical space by a lattice
$\ZZZ^3$; after all, lattice approaches have often been employed in
QFT, for various reasons. Moreover, Bell-type QFTs will usually have
as configurations the positions of a variable number of particles; so
the configuration has a certain continuous aspect, the positions, and
a certain discrete aspect, the number of particles. Sometimes one
wishes to study simplified models, and in this vein it may be
interesting to have only the particle number as a state variable, and
thus the set of nonnegative integers as configuration space.

\subsection{Bell's Process}
\label{sec:Bell}

The model Bell specified in \cite{BellBeables} is a case of a minimal
jump process on a discrete set.  ``For simplicity,'' Bell considers a
lattice $\Lambda$ instead of continuous 3-space, and a Hamiltonian of
a lattice QFT.  As a consequence, the configuration
space $\conf = \Gamma(\Lambda)$ is countable. (Bell even makes $\conf$
finite, but this is not relevant here. We also remark that according
to Bell's formulation,  even distinguishable particles have
configuration space $\Gamma (\Lambda)$.)

Bell chooses as the configuration the number of \emph{fermions} at
every lattice site, rather than the total particle number (i.e., in
our terminology he takes $\pov\{q\}$ to be the projection to the joint
eigenspace of the fermion number operators for all lattice sites with
eigenvalues the occupation numbers corresponding to $q \in
\Gamma(\Lambda)$).  He thus gives the fermionic degrees of freedom a
status different from the bosonic ones.  That is to say, boson
particles do not exist in Bell's model, despite the fact that
$\Hilbert = \Hilbert_\mathrm{fermions} \otimes
\Hilbert_\mathrm{bosons}$ and the presence of bosonic terms in the
Hamiltonian.

Thus the PVM $\pov\{q\} = \pov_\mathrm{fermions} \{q\} \otimes
\1_\mathrm{bosons}$ is ``doubly'' degenerate: the fermionic occupation
number operators do not form a complete set of commuting operators,
because of both the spin and the bosonic degrees of freedom. Different
spin states and different quantum states of the bosonic fields are
compatible with the same fermion occupation numbers.  So a further
index $i$ is necessary to label a basis $\{ |q,i\rangle \}$ of
$\Hilbert$.  The jump rates Bell prescribes are then
\eqref{degenrates}, and are thus a special case of
\eqref{tranrates}. We emphasize that here the index $i$ does not
merely label different spin states, but states of the quantized
radiation as well.

\subsection{A Case of POVM}\label{sec:expovm1}

Consider for $\Hilbert$ the space of Dirac wave functions of positive
energy. The POVM $\pov(\,\cdot\,)$ we defined on it in Section
\ref{sec:mini2} is, as we have already remarked, not a PVM but a
genuine POVM and arises from the natural PVM $\pov_0(\,\cdot\,)$ on
$L^2(\RRR^3,\CCC^4)$ by $\pov(\,\cdot\,) = P_+\pov_0(\,\cdot\,) I$
with $P_+: L^2(\RRR^3,\CCC^4) \to \Hilbert$ the projection and
$I:\Hilbert \to L^2(\RRR^3,\CCC^4)$ the inclusion mapping.  We can
extend any given interaction Hamiltonian $H$ on $\Hilbert$ to an
operator on $L^2(\RRR^3,\CCC^4)$, $H_\ext = IHP_+$.  If $H_\ext$
possesses a kernel $\sp{q}{H_\ext|q'}$, then $H$ corresponds to a jump
process, and the rates \eqref{tranrates} can be expressed in terms of
this kernel, since for $\Psi\in \Hilbert$, $\sp{\Psi}{\pov(dq) H
\pov(dq')| \Psi} = \sp{\Psi}{P_+\pov_0(dq)I H P_+\pov_0(dq')I|\Psi} =
\sp{\Psi}{\pov_0(dq) H_\ext \pov_0(dq') |\Psi} = \Psi^*(q) \,
\sp{q}{H_\ext |q'} \, \Psi(q') \, dq\, dq'$. We thus obtain
\begin{equation}\label{Diracrates}
   \sigma(q|q') = \frac{\big[ (2/\hbar)\, \Im \,\Psi^*(q) \,
   \sp{q}{H_\ext |q'} \, \Psi(q') \big]^+}{\Psi^*(q') \, \Psi(q')}.
\end{equation}

This POVM is used in the pair creation model of
\cite[Sec.~3.3]{crea2B}.

\subsection{Another Case of POVM}\label{sec:expovm2}

Let $\Hilbert = L^2(\RRR^d)$ and let $\pov_0(\,\cdot\,)$ be  the natural
PVM.
We obtain a POVM $\pov$ by smearing out $\pov_0$ with a profile
function $\profile:\RRR^d \to [0,\infty)$ with $\int \profile(q) \, dq
=1$ and $\profile(-q) = \profile(q)$, e.g., a Gaussian:
\begin{equation}\label{}
   \pov(B) = \int\limits_{q \in B} dq \int\limits_{q' \in \RRR^d}
   \profile(q'-q) \, \pov_0(dq').
\end{equation}
Whereas $\pov_0(B)$ is  multiplication by $\1_B$, $\pov(B)$ is
multiplication by $\profile \star \1_B$. It leads to $\measure(dq) =
(\profile \star |\Psi|^2)(q) \, dq$.

The jump rate formula \eqref{tranrates} then yields
\[
   \sigma(q|q') = \frac{\big[(2/\hbar) \, \Im \int dq'' \int dq''' \,
   \profile(q''-q) \Psi^*(q'') \, \sp{q''}{H|q'''} \, \Psi(q''') \,
   \profile(q'''-q') \big]^+}{\int dq'' \,\profile(q''-q')
   \,\Psi^*(q'') \,\Psi(q'')},
\]
i.e., the denominator gets smeared out with $\profile$, and the square
bracket in the numerator gets smeared out with $\profile$ in each
variable.

\subsection{Identical Particles}\label{sec:identical}

The $n$-particle sector of the configuration space (without
coincidence configurations) of identical particles
$\Gommo(\RRR^3)$ is the manifold of $n$-point subsets of
$\RRR^3$; let $\conf$ be this manifold. The most common way of
describing the quantum state of $n$ fermions is by an anti-symmetric
(square-integrable) wave function $\Psi$ on $\hat\conf := \RRR^{3n}$;
let $\Hilbert$ be the space of such functions.  Whereas for bosons
$\Psi$ could be viewed as a function on $\conf$, for fermions $\Psi$
is not a function on $\conf$.

Nonetheless, the configuration observable still corresponds to a PVM
$\pov$ on $\conf$: for $B \subseteq \conf$, we set $\pov(B)
\Psi(\vq_1, \ldots, \vq_n) = \Psi(\vq_1, \ldots, \vq_n)$ if $\{\vq_1,
\ldots, \vq_n\} \in B$ and zero otherwise. In other words, $\pov(B)$
is multiplication by the indicator function of $\covering^{-1}(B)$
where $\covering$ is the obvious projection mapping $\hat\conf
\setminus \Delta \to \conf$, with $\Delta$ the set of coincidence
configurations.

To obtain other useful expressions  for this PVM, we introduce the
formal kets $|\hat{q} \rangle$ for $\hat{q} \in \hat\conf$ (to be
treated like elements of $L^2(\hat\conf)$), the anti-symmetrization
operator $S$ (i.e., the projection $L^2(\hat\conf) \to \Hilbert$), the
normalized anti-symmetrizer\footnote{The name means this: since $S$ is
a projection, $S \Psi$ is usually not a unit vector when $\Psi$ is.
Whenever $\Psi \in L^2(\hat\conf)$ is supported by a fundamental
domain of the permutation group, i.e., by a set $\Omega \subseteq
\hat\conf$ on which (the restriction of) $\covering$  is a bijection
to $\conf$, the norm of $S\Psi$ is $1/\sqrt{n!}$, so that $s\Psi$ is
again a unit vector.} $s= \sqrt{n!} \, S$, and the formal kets $|s
\hat{q}\rangle := s|\hat{q} \rangle$ (to be treated like elements of
$\Hilbert$). The $|\hat{q} \rangle$ and $|s\hat{q} \rangle$ are
normalized in the sense that
\[
   \sp{\hat{q}} {\hat{q}'} = \delta(\hat{q} - \hat{q}') \text{ and }
   \sp{s\hat{q}} {s\hat{q}'} = (-1)^{\permutation(\hat{q},\hat{q}')} \,
   \delta(q-q'),
\]
where $q=\covering(\hat{q})$, $q'=\covering(\hat{q}')$,
$\permutation(\hat{q},\hat{q}')$ is the permutation that carries
$\hat{q}$ into $\hat{q}'$ given that $q=q'$, and $(-1)^\permutation$
is the sign of the permutation $\permutation$. Now we can write
\begin{equation}\label{idenpovm}
   \pov(dq) = \sum_{\hat{q} \in \covering^{-1}(q)} |\hat{q} \rangle
   \langle \hat{q}| \, dq = n! \, S |\hat{q} \rangle \langle
   \hat{q}| \, dq = |s\hat{q} \rangle \langle s\hat{q}| \, dq,
\end{equation}
where the sum is over the $n!$ ways   of numbering the $n$
points in $q$; the last two terms actually do not depend on the choice
of $\hat{q} \in \covering^{-1}(q)$, the numbering of $q$.

The probability distribution arising from this PVM is
\begin{equation}\label{idenmeasure}
   \measure(dq) = \sum_{\hat{q} \in \covering^{-1}(q)}
   |\Psi(\hat{q})|^2 \, dq = n! \, |\Psi(\hat{q})|^2 \, dq =
   |\sp{s\hat{q}}{\Psi}|^2 \, dq
\end{equation}
with arbitrary $\hat{q} \in \covering^{-1}(q)$.

If an operator $\hat{H}$ on $L^2(\hat\conf)$ is permutation invariant,
\begin{equation}\label{perminv}
   U_\permutation^{-1} \hat{H} U_\permutation = \hat{H} \text{ for every
   permutation } \permutation,
\end{equation}
where $U_\permutation$ is the unitary operator on $L^2(\hat\conf)$
performing the permutation $\permutation$, then $\hat{H}$ maps
anti-symmetric functions to anti-symmetric functions, and thus defines
an operator $H$ on $\Hilbert$. If $\hat{H}$ has a kernel
$\sp{\hat{q}}{\hat{H}| \hat{q}'}$ then the kernel is permutation
invariant in the sense that
\begin{equation}
   \sp{\permutation (\hat{q})}{\hat{H}| \permutation (\hat{q}') } =
   \sp{\hat{q}}{\hat{H}| \hat{q}'} \quad \forall
   \permutation,
\end{equation}
where $\permutation(\vq_1, \ldots,\vq_n) := (\vq_{\permutation(1)},
\ldots, \vq_{\permutation(n)})$, and $H$ also possesses a kernel,
\[
   \sp{s\hat{q}}{H|s\hat{q}'} = n! \, \sp{\hat{q}}{S\hat{H}S|\hat{q}'}
   =\frac{1}{n!}  \sum\limits_{\permutation,\permutation'}
   \sp{\permutation (\hat{q})}{\hat{H}| \permutation' (\hat{q}') } .
\]
In this case \eqref{tranrates} yields
\begin{subequations}\label{idenrates}
\begin{align}
   \sigma(q|q') &= \frac{\Big[\tfrac{2}{\hbar} \, \Im
   \sum\limits_{\hat{q}, \hat{q}'} \Psi^*(\hat{q}) \,
   \sp{\hat{q}}{\hat{H}| \hat{q}'} \, \Psi( \hat{q}')
   \Big]^+}{\sum\limits_{\hat{q}'} \Psi^*(\hat{q}') \,
   \Psi(\hat{q}')} \label{idenratesa}\\
   &= \frac{\Big[\tfrac{2}{\hbar} \, \Im \, \sp{\Psi}{s\hat{q}}
   \sp{s\hat{q}}{H| s\hat{q}'} \sp{s\hat{q}'}{\Psi} \Big]^+}
   {\sp{\Psi}{s\hat{q}'} \sp{s\hat{q}'}{\Psi}} \label{idenratesb}
\end{align}
\end{subequations}
where $\hat{q} \in \covering^{-1}(q)$ and $\hat{q}' \in
\covering^{-1}(q')$, as running variables in \eqref{idenratesa} and as
arbitrary but fixed in \eqref{idenratesb}.

\subsection{Another View of Fermions}

There is a way of viewing fermion wave functions as being defined on
$\conf$, rather than $\RRR^{3n}$, by regarding  them as cross-sections 
of
a particular 1-dimensional vector bundle over $\conf$. To this end,
define an $n!$-dimensional vector bundle $E$ by
\begin{equation}\label{idenEdef}
   E_q := \bigoplus_{\hat{q} \in \covering^{-1}(q)} \CCC\,.
\end{equation}
Every function $\Psi:\RRR^{3n} \to \CCC$ naturally gives rise to a
cross-section $\Phi$ of $E$, defined by
\begin{equation}
   \Phi(q) := \bigoplus_{\hat{q} \in \covering^{-1}(q)} \Psi(\hat{q})\,.
\end{equation}
The anti-symmetric functions form a 1-dimensional subbundle of $E$
(see also \cite{identical} for a discussion of this bundle).  The jump
rate formula for vector bundles \eqref{bundlerates} can be applied to
either the subbundle or $E$, depending on the way in which the kernel
of $H$ is given. The kernel $\sp{\hat{q}} {\hat{H}| \hat{q}'}$ above
translates directly into a kernel on $\conf \times \conf$ with values
in $E_q \otimes E^*_{q'}$, for which the rate formula for bundles
\eqref{bundlerates} is the same as the rate formula for identical
particles \eqref{idenratesa} derived in the previous section.

Another alternative view of a fermion wave function is to regard it as
a complex differential form of full rank, a $3n$-form, on
$\conf$. (See, e.g., \cite{identical}.  This would not work if the
dimension of physical space were even.) Of course, the complex
$3n$-forms are nothing but the sections of a certain 1-dimensional
bundle, usually denoted $\CCC\otimes \Lambda^{3n} \conf$, which is
equivalent to the subbundle of $E$ considered in the previous
paragraph, and which is contained in the bundle $\CCC \otimes \Lambda
\conf$ of Grassmann numbers over $\conf$.

\subsection{A Simple QFT}\label{sec:crea1}

We presented a simple example of a Bell-type QFT in \cite{crea1},
and we will now briefly point to the aspects of this model that are
relevant here. The model is based on one of the simplest possible QFTs
\cite[p.~339]{Schweber}.

The relevant configuration space $\conf$ for a QFT (with a single
particle species) is the configuration space of a variable number of
identical particles in $\RRR^3$, which is the set $\Gamma(\RRR^3)$,
or, ignoring the coincidence configurations (as they are exceptions),
the set $\Gommo (\RRR^3)$ of all finite subsets of $\RRR^3$. The
$n$-particle sector of this is a manifold of dimension $3n$; this
configuration space is thus a union of (disjoint) manifolds of
different dimensions. The relevant configuration space for a theory
with several particle species is the Cartesian product of several
copies of $\Gommo (\RRR^3)$.  In the model of \cite{crea1}, there are
two particle species, a fermion and a boson, and thus the
configuration space is
\begin{equation}\label{conffermionboson}
   \conf = \Gommo (\RRR^3) \times \Gommo (\RRR^3).
\end{equation}
We will denote configurations by $q=(x,y)$ with $x$ the configuration
of the fermions and $y$ the configuration of the bosons.

For simplicity, we replaced in \cite{crea1} the sectors of $\Gommo
(\RRR^3) \times \Gommo (\RRR^3)$, which are manifolds, by vector
spaces of the same dimension (by artificially numbering the
particles), and obtained the union
\begin{equation}\label{crea1conf}
   \hat{\conf} = \bigcup_{n=0}^\infty (\RRR^3)^n \times
   \bigcup_{m=0}^\infty (\RRR^3)^m \,,
\end{equation}
with $n$ the number of fermions and $m$ the number of bosons. Here,
however, we will use \eqref{conffermionboson} as the configuration
space.  In comparison with \eqref{crea1conf}, this amounts to (merely)
ignoring the numbering of the particles.

$\Hilbert$ is the tensor product of a fermion Fock space and a boson
Fock space, and thus the subspace of wave functions in
$L^2(\hat{\conf})$ that are anti-symmetric in the fermion coordinates
and symmetric in the boson coordinates. Let $S$ denote the appropriate
symmetrization operator, i.e., the projection operator
$L^2(\hat{\conf}) \to \Hilbert$, and $s$ the normalized symmetrizer
\begin{equation}\label{sdef}
   s\Psi(\vx_1, \ldots, \vx_n,\vy_1, \ldots, \vy_m) = \sqrt{n!\, m!} \,
   S\Psi(\vx_1, \ldots, \vx_n,\vy_1, \ldots, \vy_m),
\end{equation}
i.e., $s = \sqrt{N! \, M!} \, S$ with $N$ and $M$ the fermion and
boson number operators, which commute with $S$ and with each other.
As in Section \ref{sec:identical}, we denote by $\covering$ the
projection mapping $\hat{\conf} \setminus \Delta \to \conf$,
$\covering(\vx_1, \ldots, \vx_n,\vy_1, \ldots, \vy_m) = (\{\vx_1,
\ldots,\vx_n\}, \{\vy_1, \ldots, \vy_m\})$.  The configuration PVM
$\pov(B)$ on $\conf$  is multiplication by
$\1_{\covering^{-1}(B)}$, which can be understood as acting on
$\Hilbert$, though it is defined on $L^2(\hat{\conf})$, since it is
permutation invariant and thus maps $\Hilbert$ to itself.  We utilize
again the formal kets $|\hat{q}\rangle$ where $\hat{q} \in \hat{\conf}
\setminus \Delta$ is a numbered configuration, for which we also write
$\hat{q} = (\hat{x},\hat{y}) = (\vx_1, \ldots, \vx_n,\vy_1, \ldots,
\vy_m)$. We also use the symmetrized and normalized kets $|s\hat{q}
\rangle = s|\hat{q} \rangle$. As   in \eqref{idenpovm}, we can write
\begin{equation}\label{crea1povm}
   \pov(dq) = \sum_{\hat{q} \in \covering^{-1}(q)} |\hat{q} \rangle
   \langle \hat{q}| \, dq = n!\, m! \, S |\hat{q} \rangle \langle
   \hat{q}| \, dq = |s\hat{q} \rangle \langle s\hat{q}| \, dq
\end{equation}
with arbitrary $\hat{q} \in \covering^{-1}(q)$. For the probability
distribution, we thus have, as  in \eqref{idenmeasure},
\begin{equation}\label{crea1measure}
   \measure(dq) = \sum_{\hat{q} \in \covering^{-1}(q)}
   |\Psi(\hat{q})|^2 \, dq  = n!\, m! \, |\Psi(\hat{q})|^2 \, dq =
   |\sp{s\hat{q}}{\Psi}|^2 \, dq
\end{equation}
with arbitrary $\hat{q} \in \covering^{-1}(q)$.

The free Hamiltonian is the second quantized Schr\"odinger operator
(with zero potential), associated with the free process described in
Section~\ref{sec:BellQFT}.  The interaction Hamiltonian is defined by
\begin{equation}\label{HIdef}
   H_\inter = \int d^3\vx \, \psi^\dag(\vx)\, (a^\dag_\profile(\vx) +
   a_{\profile}(\vx))\, \psi(\vx)
\end{equation}
with $\psi^\dag(\vx)$ the creation operators  (in position
representation), acting on the \emph{fermion} Fock space, and
$a^\dag_\profile(\vx)$ the creation operators  (in position
representation), acting on the \emph{boson} Fock space, regularized
through convolution with an $L^2$ function $\profile:\RRR^3 \to \RRR$.
$H_\inter$ has a kernel; we will now obtain a formula for it, see
\eqref{crea1kernel} below. The $|s\hat{q} \rangle$ are connected to
the creation operators according to
\begin{equation}\label{shatqpsia}
   |s\hat{q}\rangle = \psi^\dag(\vx_n) \cdots
    \psi^\dag(\vx_1) a^\dag(\vy_m) \cdots a^\dag(\vy_1) |0\rangle\,,
\end{equation}
where $|0\rangle \in \Hilbert$ denotes the vacuum state. A relevant
fact is that the creation and annihilation operators
$\psi^\dag,\psi,a^\dag$ and $a$ possess kernels. Using the canonical
(anti\nobreakdash-)commutation relations for $\psi$ and $a$, one obtains
from
\eqref{shatqpsia} the following formulas for the kernels of
$\psi(\vr)$ and $a(\vr)$, $\vr \in \RRR^3$:
\begin{align}
   \sp{s\hat{q}}{\psi(\vr)|s\hat{q}'} &= \delta_{n,n'-1} \,
   \delta_{m,m'} \,
   \delta^{3n'}(x \cup \vr -x') \, (-1)^{\permutation((\hat{x},
   \vr),\hat{x}')} \, \delta^{3m}(y-y') \label{psikernel} \\
   \sp{s\hat{q}}{a(\vr)|s\hat{q}'} &= \delta_{n,n'} \,
   \delta_{m,m'-1} \, \delta^{3n}(x-x') \,
   (-1)^{\permutation(\hat{x},\hat{x}')} \,
   \delta^{3m'}(y \cup \vr - y') \label{akernel}
\end{align}
where $(x,y) = q = \covering(\hat{q})$, and $\permutation
(\hat{x},\hat{x}')$ denotes the permutation that carries $\hat{x}$ to
$\hat{x}'$ given that $x=x'$. The corresponding formulas for
$\psi^\dag$ and $a^\dag$ can be obtained by exchanging $\hat{q}$ and
$\hat{q}'$ on the right hand sides of \eqref{psikernel} and
\eqref{akernel}.  For the smeared-out operator $a_\profile(\vr)$, we
obtain
\begin{equation}\label{aprofilekernel}
   \sp{s\hat{q}}{a_\profile(\vr)|s\hat{q}'} = \delta_{n,n'} \,
   \delta_{m,m'-1} \, \delta^{3n}(x-x') \,
   (-1)^{\permutation(\hat{x},\hat{x}')} \sum_{\vy' \in y'}
   \delta^{3m}(y- y'\setminus \vy') \, \profile(\vy' - \vr)
\end{equation}
We make use of the resolution of the identity
\begin{equation}\label{resolution}
   I = \int\limits_{\conf} dq \, |s\hat{q} \rangle \langle
   s\hat{q}|\,.
\end{equation}
Inserting \eqref{resolution} twice into \eqref{HIdef} and exploiting
\eqref{psikernel} and \eqref{aprofilekernel}, we find
\begin{equation}\label{crea1kernel}
\begin{split}
   \sp{s\hat{q}} {H_\inter| s\hat{q}'} &= \delta_{n,n'} \,
   \delta_{m-1,m'} \, \delta^{3n}(x-x') \,
   (-1)^{\permutation(\hat{x},\hat{x}')} \sum_{\vy \in y} \delta^{3m'}
   (y \setminus \vy - y') \sum_{\vx \in x} \profile(\vy - \vx) \:  \\
   &+ \delta_{n,n'} \, \delta_{m'-1,m} \,
   \delta^{3n}(x-x') \, (-1)^{\permutation(\hat{x},\hat{x}')}
   \sum_{\vy' \in y'} \delta^{3m} (y - y' \setminus \vy') \sum_{\vx \in
   x} \profile(\vy' - \vx)\,.
\end{split}
\end{equation}
This is another case of a kernel containing $\delta$ functions (see
Section \ref{sec:ex3}).

By \eqref{crea1povm}, the jump rates \eqref{tranrates} are
\begin{equation}
   \sigma(q|q') = \frac{\Big[\tfrac{2}{\hbar} \, \Im \,
   \sp{\Psi}{s\hat{q}} \sp{s\hat{q}}{H_\inter| s\hat{q}'}
\sp{s\hat{q}'}{\Psi}
   \Big]^+} {\sp{\Psi}{s\hat{q}'} \sp{s\hat{q}'}{\Psi}} \,.
\end{equation}
More explicitly, we obtain from \eqref{crea1kernel} the rates
\begin{equation}\label{crea1rates}
\begin{split}
   \sigma(q|q') &= \delta_{nn'} \,\delta_{m-1,m'} \,\delta^{3n}(x-x')
   \sum_{\vy \in y} \delta^{3m'}(y\setminus \vy-y') \,
   \sigma_\crea(q'\cup \vy|q') \:  \\
   &+\delta_{nn'}\,\delta_{m,m'-1} \, \delta^{3n}(x-x') \sum_{\vy' \in
   y'} \delta^{3m}(y - y'\setminus \vy') \, \sigma_\ann(q'\setminus
   \vy'|q')
\end{split}
\end{equation}
with
\begin{subequations}
\begin{align}
   \sigma_\crea(q'\cup \vy|q')&= \frac{2 \sqrt{m'+1}}{\hbar} \,
   \frac{\Big[ \Im \, \Psi^*(\hat{q}) \,
   (-1)^{\permutation(\hat{x},\hat{x}')} \sum\limits_{\vx' \in x'}
   \varphi(\vy-\vx') \, \Psi(\hat{q}')\Big]^+}{ \Psi^*(\hat{q}') \,
   \Psi(\hat{q}')} \label{crea1crearate} \\
   \sigma_\ann(q'\setminus \vy'|q')&= \frac{2} {\hbar \sqrt{m'}}
   \,\frac{\Big[\Im \, \Psi^*(\hat{q}) \,
   (-1)^{\permutation(\hat{x},\hat{x}')} \sum\limits_{\vx' \in x'}
   \varphi(\vy'-\vx') \, \Psi(\hat{q}') \Big]^+}{ \Psi^*(\hat{q}') \,
   \Psi(\hat{q}')} , \label{crea1annrate}
\end{align}
\end{subequations}
for arbitrary $\hat{q}' \in \covering^{-1}(q')$ and $\hat{q} \in
\covering^{-1}(q)$ with $q=(x',y'\cup\vy)$ respectively $q=(x',y'
\setminus \vy')$.  (Note that a sum sign can be drawn out of the plus
function if the terms have disjoint supports.)

Equation \eqref{crea1rates} is worth looking at closely: One can read
off that the only possible jumps are $(x',y') \to (x',y' \cup \vy)$,
creation of a boson, and $(x',y') \to (x',y' \setminus \vy')$,
annihilation of a boson. In particular, while one particle is created
or annihilated, the other particles do not move. The process that we
considered in \cite{crea1} consists of pieces of Bohmian trajectories
interrupted by jumps with rates \eqref{crea1rates}; the process is
thus an example of the jump rate formula \eqref{tranrates}, and an
example of combining jumps and Bohmian motion by means of process
additivity.

The example shows how, for other QFTs, the jump rates
\eqref{tranrates} can be applied to relevant interaction Hamiltonians:
If $H_\inter$ is, in the position representation, a polynomial in the
creation and annihilation operators, then it possesses a kernel on the
relevant configuration space. A cut-off (implemented here by smearing
out the creation and annihilation operators) needs to be introduced to
make $H_\inter$ a well-defined operator on $L^2$.

If, in some QFT, the particle number operator is not conserved, jumps
between the sectors of configuration space are inevitable for an
equivariant process. And, indeed, when $H_\inter$ does not commute
with the particle number operator (as is usually the case), jumps can
occur that change the number of particles. Often, $H_\inter$ contains
\emph{only} off-diagonal terms with respect to the particle number;
then every jump will change the particle number.  This is precisely
what happens in the model of \cite{crea1}.

\section{Existence Results}
\label{sec:maths}

The configuration space $\conf$ is assumed in this paper to be a
measurable space, equipped with a $\sigma$-algebra $\salg$.  Every set
we consider is assumed to belong to the appropriate $\sigma$-algebra:
$\salg$ on $\conf$ or the product $\sigma$-algebra $\salg \otimes
\salg$ on $\conf \times \conf$. If $\form$ is a quadratic form, we
will usually use the notation $\sp{\Phi} {\form | \Psi}$ rather than
$\form (\Phi,\Psi)$.  If $\pov(B)\Psi$ and $\pov(C)\Psi$ lie in the
form domain of $H$, we write $\sp{\Psi}{\pov(B) H \pov(C)|\Psi}$ for
$\sp{\pov(B)\Psi}{H|\pov(C)\Psi}$.

\subsection{Condition for Finite Rates}\label{sec:thm}

For the argument of Section \ref{sec:mini2} to work, it is
necessary that (a) the bracket in the numerator of \eqref{tranrates}
exist  as a finite signed measure on $\conf \times \conf$, and (b) the
Radon--Nikod\'ym derivative of the numerator with respect to the
denominator  also be  well defined. It turns out that, given (a), (b) is
straightforward. However, contrary to what a superficial inspection
might suggest, (a) is problematical even when $H$ is bounded. To see
this, consider the case $\Hilbert = L^2(\RRR)$ with the natural PVM
(corresponding to position) on $\conf = \RRR$, and with $H$ the sum of
the Fourier transform on $\Hilbert$ and its adjoint, given by the
kernel
\[
   \sp{q}{H|q'} = \sqrt{\frac{2}{\pi}} \, \cos(qq')\,.
\]
Then, for $\Psi$ real, the bracket in \eqref{tranrates} would have to
be understood as proportional to
\[
   \Psi(q) \, \cos(qq') \, \Psi(q')\,,
\]
and $\Psi \in \Hilbert$ could be so chosen that this does not define a
signed measure on $\RRR \times \RRR$ because both its positive and
negative part have infinite total weight. In fact, $\Psi$ can be so
chosen that the resulting $\sigma(\,\cdot\,|q')$ is an infinite
measure, $\sigma(\conf|q') = \infty$, for all $q'$, and thus does not
define a jump process. Note, however, that for $\Psi \in L^1(\RRR)
\cap L^2(\RRR)$, $\sigma(\,\cdot\,|q')$ is finite for this $H$.

The following theorem provides a condition under which the argument
sketched in Section \ref{sec:mini2} for the equivariance of the jump
rates $\sigma$, steps 1--5, can be made rigorous.

\begin{theorem}\label{sta:finiterates}
   Let $\Hilbert$ be a Hilbert space, $\Psi\in \Hilbert$ with
   $\|\Psi\|=1$, $H$ a self-adjoint operator on $\Hilbert$, $\conf$ a
   standard Borel space,\footnote{A \emph{standard Borel space} is a
   measurable space isomorphic to a complete separable metric space
   with its Borel $\sigma$-algebra.  Basically all spaces that arise
   in practice are in fact standard Borel spaces, and so are in
   particular all spaces that we have in mind for $\conf$ (which are
   countable unions of (separable) Riemannian manifolds). Thus, the
   condition of being a standard Borel space is not much of a
   restriction.} and $\pov$ a POVM on $\conf$ acting on $\Hilbert$.
   Suppose that for all $B \subseteq \conf$, $\pov(B) \Psi$ lies in
   the form domain of $H$, and there exists a complex measure $\mu$ on
   $\conf \times \conf$ such that for all $B,C \subseteq \conf$,
   \begin{equation}\label{muPsiPH}
     \mu(B \times C) = \sp{\Psi}{\pov(B) H \pov(C)| \Psi}\,.
   \end{equation}
   Then the jump rates \eqref{tranrates} are well-defined and finite
   for $\measure$-almost every $q'$, and they are equivariant if, in
   addition, $\Psi \in \mathrm{domain}(H)$.
\end{theorem}

\begin{proof}
We first show that under the hypotheses of the theorem, the jump rates
(\ref{tranrates}) are well-defined and finite.  Then we show that they
are equivariant.

To begin with, the measure $\mu$ whose existence was assumed in the
theorem is conjugate symmetric under the transposition mapping
$(q,q')^\tr = (q',q)$ on $\conf \times \conf$, i.e., $\mu(A^\tr) =
\mu(A)^*$. To see this, note that a complex measure on $\conf \times
\conf$ is uniquely determined by its values on product
sets. $\mu(\,\cdot\,^\tr)$ and $\mu(\,\cdot\,)^*$ must thus be the
same measure since, by the self-adjointness of $H$, for $A = B \times
C$, $\mu(A^\tr) = \mu(C \times B) \stackrel{\eqref{muPsiPH}}{=}
\sp{\Psi}{\pov(C) H \pov(B) |\Psi} = \sp{\Psi}{\pov(B) H \pov(C)
|\Psi}^* = \mu(A)^*$.

We define a signed measure $\current$ on $\conf \times \conf$ by
$\current = \frac{2}{\hbar} \, \Im \, \mu$. Let $\current^+$ be the
positive part of $\current$ (defined by its Hahn--Jordan
decomposition, $\current = \current^+ - \current^-$, see e.g.\
\cite[p.~120]{Halmos}). Since $\mu$ is a complex measure (and thus
assumes only finite values), $\current$ has finite positive and
negative parts. Since $\mu$ is conjugate symmetric, $\current$ is
anti-symmetric.

We now show that for every $B\subseteq \conf$, the measure
$\current(B\times \,\cdot\,)$ on $\conf$ is absolutely continuous with
respect to $\measure(\,\cdot\,)$, the ``$|\Psi|^2$'' measure defined
in (\ref{mis}). If $C$ is a $\measure$-null set, that is $\sp{\Psi}
{\pov(C)| \Psi}=0$, then $\pov(C)|\Psi\rangle =0$: if $\pov(C)$ is a
projection, this is immediate, and if $\pov(C)$ is just any positive
operator, it follows from the spectral theorem---any component of
$\Psi$ orthogonal to the eigenspace of $\pov(C)$ with eigenvalue zero
would lie in the positive spectral subspace of $\pov(C)$ and give a
positive contribution to $\sp{\Psi}{\pov(C)| \Psi}$. {}From $\pov(C)
\Psi =0$ it follows that $\sp{ \Psi}{\pov(B) H \pov(C) |\Psi} =0$, so
that $\current(B \times C) =0$, which is what we wanted to show.

Next we show that for every $B \subseteq \conf$, the measure
$\current^+(B \times \,\cdot\,)$ is absolutely continuous with respect
to $\measure(\,\cdot\,)$. Suppose again that $\measure(C)=0$.  We have
that
\[
  \current^+(B\times C) \leq \current^+(B\times C) +
  \current^-(B\times C)=
\]
\[
  = |\current|(B\times C) = \sup \sum_{i,j} |\current(B_i\times C_j)|
\]
where the sup is taken over all finite partitions $\bigcup_i B_i = B$
of $B$ and $\bigcup_j C_j = C$ of $C$. Now each $\current(B_i\times
C_j) =0$ because $\current(B_i \times \,\cdot\,) \ll
\measure(\,\cdot\,)$ and $\measure(C_j) \leq \measure(C) =0$. Thus
$\current^+(B\times C) = 0$.

It follows from the Radon--Nikod\'ym theorem that for every $B$,
$\current^+(B\times\,\cdot\,)$ possesses a density with respect to
$\measure(\,\cdot\,)$.  The density is unique up to changes on
$\measure$-null sets, and one version of this density is what we will
take as $\sigma(B|q')$.  We have to make sure, though, that $\sigma$
is a measure in its dependence on $B$, and from the Radon--Nikod\'ym
theorem alone we do not obtain additivity in $B$. For this reason, we
utilize a standard theorem \cite[p.~147]{Partha} on the existence of
regular conditional probabilities, asserting that if $\conf$ (and thus
also $\conf \times \conf$) is a standard Borel space, then every
probability measure $\nu$ on $\conf \times \conf$ possesses regular
conditional probabilities, i.e., a function $p(\,\cdot\,|q')$ on
$\conf$ with values in the probability measures on $\conf \times
\conf$ such that for almost every $q'$, $p(\,\cdot\,|q')$ is
concentrated on the set $\conf \times \{q'\} \subseteq \conf \times
\conf$, and for every $A\subseteq \conf \times \conf$, $p(A|q')$ is a
measurable function of $q'$ with
\begin{equation}\label{regcondprob}
   \int\limits_{q' \in \conf} p(A|q') \, \nu(\conf \times dq') =
   \nu(A).
\end{equation}
We set $\nu(\,\cdot\,) = \current^+(\,\cdot\,)/\current^+(\conf \times
\conf)$ and define $\sigma$ as the corresponding regular conditional
probability times a factor that takes into account that
\eqref{tranrates} involves the density of $\current^+$ relative to
$\measure$ (rather than to $\nu(\conf \times \,\cdot\,)$ or
$\current^+(\conf \times \,\cdot\,)$):
\begin{equation}\label{sigmadef}
   \sigma(B|q') := p(B \times \conf|q') \, \frac{d\current^+(\conf
   \times \,\cdot\,)} {d\measure(\,\cdot\,)}(q') \,.
\end{equation}
The last factor exists because we have shown above that
$\current^+(\conf \times \,\cdot\,) \ll \measure(\,\cdot\,)$.
$\sigma(\,\cdot\,|q')$ is a (finite) measure because $p(\,\cdot\,|q')$
is.  For fixed $B$, $\sigma(B|q')$ as a function of $q'$ is a version
of the Radon--Nikod\'ym derivative $d\current^+(B \times
\,\cdot\,)/d\measure(\,\cdot\,)$ because
\[
   \int\limits_{q' \in C} \sigma(B|q') \, \measure(dq')
   \stackrel{\eqref{sigmadef}}{=} \int\limits_{q' \in C} p(B \times
   \conf|q') \, \frac{d\current^+(\conf \times \,\cdot\,)} {d\measure
   (\,\cdot\,)}(q') \, \measure(dq') =
\]
\[
   = \current^+(\conf \times \conf) \int\limits_{q' \in \conf} p(B
   \times C|q') \,\frac{\current^+(\conf \times dq')}{\current^+(\conf
   \times \conf)} \stackrel{\eqref{regcondprob}}{=} \current^+(B \times
   C).
\]
According to the theorem on regular conditional probabilities that we
used, $\sigma$ is defined uniquely up to changes on a $\measure$-null
set of $q'$s.

Now we check the  equivariance of the jump rates $\sigma$: for any
$B\subseteq \conf$,
\[
   \generator_\sigma \measure (B) \stackrel{\eqref{continuity3}}{=}
   \int\limits_{q'\in\conf} \sigma(B|q') \, \measure(dq') -
   \int\limits_{q\in B} \sigma(\conf|q) \, \measure(dq) =
   \current^+(B\times\conf) - \current^+(\conf\times B)\,,
\]
using that $\sigma$ is a version of the Radon--Nikod\'ym derivative of
$\current^+$ relative to $\measure$. Since $\current$ is anti-symmetric
with respect to the permutation mapping $(q,q') \mapsto (q',q)$ on
$\conf \times \conf$, we have that  $\current^+(C \times B) = \current^-
(B
\times C)$, and therefore
\[
   \generator_\sigma \measure (B) = \current^+(B\times\conf) -
   \current^-(B\times \conf) = \current(B\times\conf) =
\]
\[
   = \tfrac{2}{\hbar} \, \Im \, \mu(B \times \conf)
   \stackrel{\eqref{muPsiPH}}{=} \tfrac{2}{\hbar} \, \Im \, \sp{\Psi}
   {\pov(B) H |\Psi}.
\]

  It remains to be shown that $\measure_t(B) = \sp{\E^{-\I Ht/\hbar}
\Psi}{\pov(B) \, \E^{-\I Ht/\hbar} \Psi}$ is differentiable with
respect to time at $t=0$ and has derivative
\begin{equation}
   \frac{d\measure_t(B)}{dt} \Big|_{t=0} = \tfrac{2}{\hbar} \, \Im \,
   \sp{\Psi}{\pov(B) H|\Psi}.
\end{equation}
If $\Psi$ lies in the domain of $H$, $\Psi_t = \E^{-\I Ht/\hbar} \Psi$
is differentiable with respect to $t$ at $t=0$ \cite[p.~265]{RS} and
has derivative $\dot{\Psi} = -\frac{\I}{\hbar} H\Psi$. Hence
\begin{align}
   &\frac{1}{t} \Big( \sp{\Psi_t}{\pov(B) \Psi_t}
   - \sp{\Psi_0}{\pov(B) \Psi_0} \Big) \nonumber \\
   =&
   \sp{\Psi_t}{ \pov(B) |(\Psi_t -\Psi_0)/\,t\,} +
   \sp{(\Psi_t -\Psi_0)/\,t\,}{ \pov(B) \Psi_0}  \nonumber
\end{align}
converges, as $t \to 0$, to
\[
   \sp{\Psi}{\pov(B) \dot{\Psi}} + \sp{\dot{\Psi}} {\pov(B) \Psi} =
   -\tfrac{\I}{\hbar} \sp{\Psi}{\pov(B) H \Psi} + \tfrac{\I}{\hbar}
   \sp{H\Psi}{\pov(B) \Psi} = \tfrac{2}{\hbar} \Im \sp{\Psi}{\pov(B) H
   \Psi}.
\]
It now follows that $\generator_\sigma \measure = d\measure/dt$, which
completes the proof.
\end{proof}

We remark that if, as supposed in Theorem \ref{sta:finiterates}, the
measure $\mu$ exists, it is also unique. This follows from the fact,
which we have already mentioned, that a (complex) measure on $\conf
\times \conf$ is uniquely determined by its values on the product sets
$B \times C$.

Another remark concerns how the (existence)  assumption of Theorem
\ref{sta:finiterates} can be violated.  Since the example Hamiltonian
of Section \ref{sec:inftyrates} leads to infinite jump rates, it also
provides an example for which the assumption of Theorem
\ref{sta:finiterates} is violated, in fact  for \emph{every} nonzero
$\Psi \in \Hilbert$. To see this directly, note that, while $\pov(B)
\Psi$ lies indeed in the form domain of $H$ (which is $\Hilbert$ since 
$H$
is bounded),
\[
   \sp{\Psi}{\pov(B) H \pov(C) |\Psi} = \I \int\limits_{B} dq \:
   \text{P-}\!\int\limits_{C} dq' \, \frac{\Psi^*(q) \, \Psi(q')}{q-q'}
\]
where $\text{P-}\int$ denotes a principal value integral. For $B \cap
C = \emptyset$, $\text{P-}\int$ can be replaced by a Lebesgue
integral. This, together with \eqref{muPsiPH}, would leave for $\mu$
only one possibility (up to addition of a complex measure concentrated
on the diagonal $\{(q,q): q \in \conf\}$), namely
\[
   \mu(dq \times dq') = \I \frac{\Psi^*(q) \, \Psi(q')}{q-q'}\, dq\, dq'.
\]
But this is not a complex measure for any $\Psi$ since $\I \Psi^*(q)
\, \Psi(q')/(q-q')$ is not absolutely integrable.  This example also
nicely illustrates the difference between a complex bi-measure
$\nu(B,C)$, i.e., a complex measure in each variable, and a complex
measure $\mu(\,\cdot\,)$ on $\conf \times \conf$: $\sp{\Psi} {\pov(B)
H \pov(C) |\Psi}$ is here a complex bi-measure and thus defines a
finite-valued additive set function on the family of finite unions of
product sets $B \times C \subseteq \conf \times \conf$, which,
however, cannot be suitably extended to all sets $A \subseteq \conf
\times \conf$. The essential reason is that the positive and the
negative singularity in $1/(q-q')$ cancel (thanks to the use of
principal value integrals) for every product set but do not for some
nonproduct sets such as $\{(q,q'): q>q'\}$. In contrast, a (finite)
non-negative bi-measure can always be extended to a (finite) measure
on the product space; see Section \ref{sec:tensorpovm}.

A related remark on the need for the existence assumption of Theorem
\ref{sta:finiterates}. One might well have imagined that the complex
measure $\mu$ on $\conf \times \conf$, extending \eqref{muPsiPH} from
product sets, can always be constructed, at least when $H$ is bounded,
as the quantum expected value of the bounded-operator-valued measure
(BOVM) $\pov\times_{H} \pov$ on $\conf \times \conf$, the
``$H$-twisted product measure'' $\pov(dq)H\pov(dq')$ of the POVM
$\pov$ with itself---or, equivalently, the product of the POVM
$\pov(dq)$ and the BOVM $H\pov(dq')$. Indeed, the nonexistence of
$\mu$ for the Hamiltonian in the principal-value example that we have
just discussed, as well as for the Hamiltonian in the
Fourier-transform example at the beginning of this section, implies
that $\pov\times_{H}\pov$ does not exist as a BOVM in these cases; if
it did, so would $\mu$, for all $\Psi$. The Fourier-transform example
can also easily be adapted to show that the product $\pov_1\times
\pov_2$ of two POVMs need not exist as a BOVM, and in fact does not
exist when $\pov_1$ and $\pov_2$ are the most familiar PVMs for
quantum mechanics, corresponding respectively to position and
momentum. There is, however, an important special case for which the
product $\pov_1\times \pov_2$ of two POVMs does exist, in fact as a
POVM, namely when $\pov_1$ and $\pov_2$ mutually commute, i.e., when
$[\pov_1(B),\pov_2(C)]=0$ for all $B$ and $C$. This will be discussed
in Section \ref{sec:tensorpovm}.

\subsection{Integral Operators}
\label{sec:jumps}

In this section we make precise the statement that Hamiltonians with
(sufficiently regular) kernels lead to finite jump rates. In
particular, we specify a set of wave functions, depending on $H,$ that
lead to finite jump rates.

\subsubsection{Hilbert--Schmidt Operators}\label{sec:HS}

We begin with the simple case in which $\Psi$ is a complex-valued wave
function on $\conf$, so that the natural configuration POVM
$\pov(\,\cdot\,)$ is a ``nondegenerate'' PVM. What first comes to mind
as a class of Hamiltonians possessing a kernel is the class of
Hilbert--Schmidt operators; for these, the kernels are in fact
square-integrable functions on $\conf \times \conf$.

\begin{corollary}\label{sta:HSQC}
   Let $\conf$ be a standard Borel space, $\Hilbert = L^2 (\conf,
   \CCC, dq)$ with respect to a $\sigma$-finite nonnegative measure on
   $\conf$ that we simply denote $dq$, let $\Psi \in \Hilbert$ with
   $\|\Psi\| =1$, let $H$ be a self-adjoint operator on $\Hilbert$,
   and let $\pov$ be the natural PVM on $\conf$ (multiplication by
   indicator functions) acting on $L^2 (\conf, \CCC, dq)$. Suppose
   that $H$ is a Hilbert--Schmidt operator. Then, by virtue of Theorem
   \ref{sta:finiterates}, the jump rates given by \eqref{tranrates}
   are well-defined and finite $\measure$-almost everywhere, and
   equivariant. In fact, the jump rates are given by \eqref{mini1}
   with $\sp{q}{H|q'}$ the kernel function of $H$.
\end{corollary}

\begin{proof}
Since $H$ is a Hilbert--Schmidt operator, it possesses an integral
kernel $K(q,q')$ that is a square-integrable function
\cite[p.~210]{RS}, i.e., there is a function $K \in L^2(\conf \times
\conf, \CCC, dq \, dq')$ such that for all $\Phi \in \Hilbert$,
\[
   H\Phi(q) = \int\limits_{\conf} K(q,q') \, \Phi(q') \, dq'\,.
\]
Thus, for all $\Phi, \Phi' \in \Hilbert$,
\[
   \sp{\Phi}{H|\Phi'} = \int\limits_{\conf} dq \int\limits_{\conf} dq'
   \, \Phi^*(q) \, K(q,q') \, \Phi'(q') =
\]
(by Fubini's theorem, because the integrand is absolutely integrable)
\[
   = \int\limits_{\conf \times \conf} dq \, dq' \, \Phi^*(q) \, K(q,q')
   \, \Phi'(q').
\]
It follows that
\begin{subequations}\label{HSint}
\begin{align}
   \sp{\Psi}{\pov(B) H \pov(C) |\Psi} &= \int\limits_{\conf \times
   \conf} dq \, dq' \, \1_B(q) \, \Psi^*(q) \, K(q,q') \, \1_C(q') \,
   \Psi(q') =\\
   &= \int\limits_{B \times C} dq \, dq' \, \Psi^*(q) \,
   K(q,q') \, \Psi(q').
\end{align}
\end{subequations}
Note that since $H$ is bounded, its form domain is $\Hilbert$ and thus
contains all $\pov(B) \Psi$. For $A \subseteq \conf \times \conf$,
define
\[
   \mu(A) = \int\limits_A \Psi^*(q) \, K(q,q') \, \Psi(q') \, dq \, dq'
   \,.
\]
Since
\[
   \int\limits_{\conf \times \conf} |\Psi(q)| \, |K(q,q')| \,
   |\Psi(q')| \, dq \, dq' < \infty \,,
\]
$\mu(A)$ is always finite, and thus a complex measure.  \eqref{HSint}
entails that \eqref{muPsiPH} is satisfied, so that Theorem
\ref{sta:finiterates} applies.
\end{proof}

We have already remarked that every Hamiltonian $H$ can be
approximated by Hilbert--Schmidt operators $H_n$. In this context, it
is interesting to note that if $H$ is itself a Hilbert--Schmidt
operator, and if the $H_n$ converge to $H$ in the Hilbert--Schmidt
norm, then the rates $\sigma^{\Psi,H_n}$ converge to $\sigma^{\Psi,H}$
in the sense that
\[
   \int\limits_{\conf \times \conf}\big|\sigma^{\Psi,H_n}(dq|q') -
   \sigma^{\Psi,H}(dq|q')\big| \, |\Psi(q')|^2\, dq' \stackrel{n \to
   \infty}{\longrightarrow} 0.
\]

\subsubsection{Complex-Valued Wave Functions}\label{sec:Cwf}

In addition to the case of  Hilbert--Schmidt operators, Theorem
\ref{sta:finiterates}
applies in many other cases, in which the kernel $K(q,q')$ is not
square-integrable, nor  even a function but instead a measure $K(dq
\times dq')$. More precisely, $K(dq \times dq')$ should be a
\emph{$\sigma$-finite complex measure}, i.e., a product of a
complex-valued measurable function $\conf \times \conf \to \CCC$ and a
$\sigma$-finite nonnegative measure on $\conf \times \conf$. (Note
that this  terminology involves a slight abuse of language  since a
$\sigma$-finite complex measure need not be a complex measure.) The
complex measure $\mu$ assumed to exist  in Theorem \ref{sta:finiterates}
is then
\begin{equation}\label{muPsiK}
   \mu( dq \times dq') = \Psi^*(q) \, K(dq \times dq') \, \Psi(q') \,.
\end{equation}
This equation suggests that the minimal amount of regularity that we
need to assume on the kernel of $H$ is that it be a $\sigma$-finite
complex measure. Otherwise, there would be no hope that \eqref{muPsiK}
could be a complex measure for a generic wave function $\Psi$, that
vanishes at most on a set of measure $0$.  The exact conditions that
we need for applying Theorem \ref{sta:finiterates} to a Hamiltonian
$H$ with kernel $K(dq \times dq')$ are listed in the following
statement:

\begin{corollary}\label{sta:Ltwokernel}
   Let $\conf$ be a standard Borel space, $\Hilbert = L^2 (\conf,
   \CCC, dq)$ with respect to a $\sigma$-finite nonnegative measure on
   $\conf$ that we simply denote $dq$, let $\Psi \in \Hilbert$ with
   $\|\Psi\| =1$, let $H$ be a self-adjoint operator on $\Hilbert$,
   and let $\pov$ be the natural PVM on $\conf$ acting on $L^2 (\conf,
   \CCC, dq)$. Suppose that $H$ has a kernel $K(dq \times dq')$ for
   $\Psi$; i.e., suppose that $K(dq \times dq')$ is a $\sigma$-finite
   complex measure on $\conf \times \conf$, and that some
   everywhere-defined version $\Psi:\conf \to \CCC$ of the
   almost-everywhere-defined function $\Psi \in L^2(\conf,\CCC,dq)$
   satisfies 
   \begin{subequations}\label{KHPsi}
    \begin{align} &
     \int\limits_{\conf \times \conf} |\Psi(q)| \, |K(dq \times dq')|
     \, |\Psi(q')| < \infty \label{KHPsifinite}\\ & \pov(B) \Psi \in
     \mathrm{form \; domain}(H) \quad \forall B \subseteq \conf
     \label{KHPsiformdomain}\\
     & \sp{\Psi}{\pov(B) H \pov(C) |\Psi} = \int\limits_{B \times C}
     \Psi^*(q) \, K(dq \times dq') \, \Psi(q') \quad \forall B,C
     \subseteq \conf. \label{KHPsiformkernel} 
    \end{align}
   \end{subequations}
   Then, by virtue of Theorem~\ref{sta:finiterates}, the jump rates
   given by \eqref{tranrates} are well-defined and finite
   $\measure$-almost everywhere, and they are equivariant if $\Psi \in
   \mathrm{domain}(H)$.
\end{corollary}

\begin{proof}
Set
\begin{equation}\label{mudef}
   \mu(A) = \int\limits_{A} \Psi^*(q) \, K(dq \times dq') \,
   \Psi(q')\,.
\end{equation}
The integral exists because of \eqref{KHPsifinite} and defines a
complex measure $\mu$, which satisfies \eqref{muPsiPH} because of
\eqref{KHPsiformkernel}.
\end{proof}

We remark that the choice of the everywhere-defined version $\Psi:
\conf \to \CCC$ of the almost-everywhere-defined function $\Psi \in
L^2(\conf, \CCC, dq)$ does not affect the jump rates, since the
measure $\mu$ is uniquely determined by its values on product sets,
which are given in \eqref{muPsiPH} in terms of the
almost-everywhere-defined function $\Psi \in \Hilbert$.

The reader may be surprised that our notion of $H$ having a kernel $K$
seems to depend on $\Psi$, whereas one may expect that $H$ either has
a kernel or does not, independent of $\Psi$. The reason for our
putting it this way is that domain questions are very delicate for
such general kernels, and it is a tricky question for which $\Psi$'s
the expression $\sp{\Psi}{\pov(B) H \pov(C) |\Psi}$ is actually given
by the integral \eqref{KHPsiformkernel}. A discussion of domain
questions would only obscure what is actually relevant for having a
situation in which Theorem \ref{sta:finiterates} applies, which is
\eqref{KHPsi}. Note, though, that if $H$ has kernel $K(dq \times dq')$
for $\Psi$, then it has kernel $K$ also for every $\Psi'$ from the
subspace spanned by $\pov(B) \Psi$ for all $B \subseteq \conf$.

The conditions \eqref{KHPsi} become very transparent in the following
case: Suppose $H$ is a self-adjoint extension of the integral operator
$K$ arising from a kernel $K(q,dq')$ that is a $\sigma$-finite complex
measure on $\conf$ for every $q\in \conf$ and is such that for every
$B \subseteq \conf$, $K(q,B)$ is a measurable function of $q$. $K$ is
defined by
\begin{equation}\label{KK}
   K\Phi(q) = \int\limits_{q' \in\conf} K(q,dq') \, \Phi(q')
\end{equation}
on the domain $\domain$ containing the $\Phi$'s satisfying
\begin{equation}
   \int\limits_{q' \in \conf} |K(q,dq')| \, |\Phi(q')| < \infty \text{ 
for
   almost every } q
\end{equation}
and
\begin{equation}
   \int\limits_\conf K(q,dq') \, \Phi(q') \text{ is an $L^2$ function of
   }q.
\end{equation}
That $H$ is an extension of $K$ means that the domain of $H$ contains
$\domain$, and $H\Phi = K\Phi$ for all $\Phi \in \domain$. Then, for a
$\Psi \in \domain$ satisfying
\begin{equation}
   \int\limits_{q' \in B} K(q,dq') \, \Psi(q') \in L^2(\conf, \CCC, dq)
   \quad \forall B \subseteq \conf
\end{equation}
and
\begin{equation}
   \int\limits_{\conf \times \conf} |\Psi(q)| \, |K(q,dq')| \,
   |\Psi(q')| \, dq < \infty,
\end{equation}
conditions \eqref{KHPsi} are satisfied with $K(dq \times dq') =
K(q,dq') \, dq$, and thus Corollary \ref{sta:Ltwokernel} applies.  The
jump rates \eqref{tranrates} can still be written as in \eqref{mini1},
understood as a measure in $q$.

Corollary \ref{sta:Ltwokernel} defines a set  of good $\Psi$'s, for
which the jump rates are finite, for the examples of Sections
\ref{sec:ex1}, \ref{sec:ex3}, and for  \eqref{disrates1}.

\subsubsection{Vector-Valued Wave Functions}

We now consider wave functions with spin, i.e., with values in
$\CCC^k$. In this case, let $\Psi^*(q)$ denote, as before, the adjoint
spinor, and $\Phi^*(q)\, \Psi(q)$ the inner product in $\CCC^k$.
Corollary \ref{sta:Ltwokernel} remains true if we replace $\CCC$ by
$\CCC^k$ everywhere and understand $K(dq \times dq')$ as
matrix-valued, i.e., as the product of a matrix-valued function and a
$\sigma$-finite nonnegative measure. The proof goes through without
changes.

Let us now be  a bit more general and allow the value space
of the wave function to vary with $q$; we reformulate Corollary
\ref{sta:Ltwokernel} for wave functions that are cross-sections of a
vector bundle $E$ over $\conf$. The kernel is then matrix valued in
the sense that $\sp{q}{H|q'}$ is a linear mapping $E_{q'} \to E_q$.

\begin{corollary}\label{sta:bundlekernel}
   Let $\conf = \bigcup_n \conf^{(n)}$ be an (at most) countable union
   of (separable) Riemannian manifolds, and $E = \bigcup_n E^{(n)}$
   the union of vector bundles $E^{(n)}$ over $\conf^{(n)}$, where the
   fiber spaces $E_q$ are endowed with Hermitian inner products, which
   we denote by $\Phi^*(q) \, \Psi(q)$. Let $\Hilbert = L^2(E,dq)$ be
   the space of square-integrable (with respect to the Riemannian
   volume measure that we denote $dq$) cross-sections of $E$, let
   $\Psi \in \Hilbert$ with $\|\Psi\| =1$, let $H$ be a self-adjoint
   operator on $\Hilbert$, and let $\pov$ be the natural PVM on
   $\conf$ acting on $\Hilbert$.  Suppose that $K(dq \times dq')$, the
   product of a $\sigma$-finite nonnegative measure on $\conf \times
   \conf$ and a section of the bundle $\bigcup_{q,q'} E_q \otimes
   E^*_{q'}$ over $\conf \times \conf$, is a kernel of $H$ for $\Psi$;
   i.e., suppose that some everywhere-defined version $\Psi$ of the
   almost-everywhere-defined cross-section $\Psi \in L^2(E,dq)$
   satisfies \eqref{KHPsifinite}-\eqref{KHPsiformkernel} (where the
   integrand on the right hand side of \eqref{KHPsiformkernel} should
   now be understood as involving the Hermitian inner product of
   $E_q$, and \eqref{KHPsifinite} as involving the operator norm of
   $K(dq \times dq')$).  Then, by virtue of Theorem
   \ref{sta:finiterates}, the jump rates given by \eqref{tranrates}
   are well-defined and finite $\measure$-almost everywhere, and they
   are equivariant if $\Psi \in \mathrm{domain}(H)$.
\end{corollary}

The proof of Corollary \ref{sta:Ltwokernel} applies here without
changes.   \eqref{mini1} remains  valid if suitably interpreted.
Corollary \ref{sta:bundlekernel} defines a set  of good $\Psi$'s, for
which the jump rates are finite, for the examples of Sections
\ref{sec:exspin}, \ref{sec:exbundle}, \ref{sec:identical},
\ref{sec:crea1}, and for \eqref{degenrates} in case the sum
over $i$ is always finite.

\subsubsection{POVMs}
\label{sec:povmkernel}

We now proceed to the fully general case of an arbitrary POVM.  First,
we provide two important mathematical tools for dealing with POVMs.

\begin{itemize}
\item Any POVM corresponds to a PVM on a larger Hilbert space,
   according to the following theorem of Naimark \cite[p.~142]{Davies}:
   \textit{If $\pov$ is a POVM on the standard Borel space $\conf$
   acting on the Hilbert space $\Hilbert$, then there is a Hilbert
   space $\Hilbert_\ext \supseteq \Hilbert$ and a PVM $\pov_\ext$ on
   $\conf$ acting on $\Hilbert_\ext$ such that $\pov(\,\cdot\,) = P_+
   \pov_\ext(\,\cdot\,) I$ with $P_+: \Hilbert_\ext \to \Hilbert$ the
   projection and $I:\Hilbert \to \Hilbert_\ext$ the inclusion, and
   $\Hilbert_\ext$ is the closed linear hull of $\{\pov_\ext(B)
   \Hilbert: B \subseteq \conf\}$. The pair $\Hilbert_\ext, \pov_\ext$
   is unique in the sense that if $\Hilbert_\ext'$, $\pov_\ext'$ is
   another such pair then there is a unitary isomorphism between
   $\Hilbert_\ext$ and $\Hilbert_\ext'$ fixing $\Hilbert$ and carrying
   $\pov_\ext$ to $\pov_\ext'$.}

   We call $\Hilbert_\ext$ and $\pov_\ext$ the \emph{Naimark extension}
   of $\Hilbert$ and $\pov$. We recall that for the Hilbert space of
   positive energy solutions of the Dirac equation and the
   corresponding POVM introduced earlier, the Naimark extension is
   given by $L^2(\RRR^3,\CCC^4)$ and its natural PVM; this example
   indicates that the Naimark extension may be, in practice, something
   natural to consider.

\item In Corollaries \ref{sta:Ltwokernel} and \ref{sta:bundlekernel},
   we were considering, for $\Hilbert$ and $\pov$, $L^2$ spaces with
   their \emph{natural} PVMs. But when we are given an \emph{arbitrary}
   PVM on a Hilbert space, the situation is not genuinely more general,
   since it can be viewed as the natural PVM of an $L^2$ space. We call
   this the \emph{naturalization} of the PVM. It is based on the
   following version of the spectral theorem (which can be obtained
   from the representation theory of abelian operator algebras, see,
   e.g., \cite{Dixmier}): \textit{If $\pov$ is a PVM on the standard
   Borel space $\conf$ acting on the Hilbert space $\Hilbert$, then
   there is a measurable field of Hilbert spaces\footnote{A
   \emph{measurable field of Hilbert spaces} on $\conf$ is a family of
   Hilbert spaces $\Hilbert_q$ with scalar products
   $\scalar{\,\cdot\,}{\,\cdot\,}_q$, endowed with a measurable
   structure that can be defined by specifying a family of
   cross-sections $\Phi_i(q)$ such that for all $i,i'$ the functions $q
   \mapsto \scalar{\Phi_i(q)}{\Phi_{i'}(q)}_q$ are measurable and for
   every $q$ the family $\Phi_i(q)$ is total in $\Hilbert_q$
   \cite{Guichardet}.}  $\Hilbert_q$ over $\conf$, a $\sigma$-finite
   nonnegative measure $dq$ on $\conf$, and a unitary isomorphism $U:
   \Hilbert \to \int^\oplus \Hilbert_q \, dq$ to the direct
   integral\footnote{This is the Hilbert space of square-integrable
   measurable cross-sections of the field $\{\Hilbert_q\}$, i.e.,
   cross-sections $\Phi(q)$ such that all functions $q \mapsto
   \scalar{\Phi_i(q)}{\Phi(q)}_q$  are measurable and $\int
   \scalar{\Phi(q)}{\Phi(q)}_q \, dq < \infty$ \cite{Guichardet}.} of
   $\Hilbert_q$ that carries $\pov$ to the natural PVM on $\conf$
   acting on $\int^\oplus \Hilbert_q \, dq$. The naturalization is
   unique in the sense that if $\{\Hilbert'_q\}, (dq)', U'$ is another
   such triple, then there is a measurable function $f:\conf \to
   (0,\infty)$ such that $(dq)' = f(q) \, dq$ and a measurable field of
   unitary isomorphisms $U_q: \Hilbert_q \to \Hilbert_q'$ such that
   $U'\Psi(q) = f(q)^{-1/2} U_q U\Psi(q)$.}

   A naturalized PVM is similar to a vector bundle in that with every
   $q\in \conf$ there is associated a value space $\Hilbert_q$, which
   however may be infinite-dimensional, and $\Psi \in \Hilbert$ can be
   understood as a function on $\conf$ such that $\Psi(q) \in
   \Hilbert_q$. Of course, instead of the differentiable structure of a
   vector bundle the naturalization of a PVM leads merely to  a
   measurable structure.
\end{itemize}

Thus, the situation with a general POVM is not much different from the
situation with a vector bundle, as treated in Corollary
\ref{sta:bundlekernel}.

For Hilbert--Schmidt operators, the kernel is so well-behaved that no
further conditions on $\Psi$ are necessary:

\begin{corollary}\label{sta:HSPOVM}
   Let $\Hilbert$ be a Hilbert space, $\Psi \in \Hilbert$ with
   $\|\Psi\| =1$, $H$ a self-adjoint operator on $\Hilbert$, $\conf$ a
   standard Borel space, and let $\pov$ be a POVM on $\conf$ acting on
   $\Hilbert$. Suppose that $H$ is a Hilbert--Schmidt operator. Then,
   by virtue of Theorem \ref{sta:finiterates}, the jump rates given by
   \eqref{tranrates} are well-defined and finite $\measure$-almost
   everywhere, and they are  equivariant.
\end{corollary}

\begin{proof}
Let $\pov_\ext$ be the Naimark extension PVM of $\pov$ acting on
$\Hilbert_\ext \supseteq \Hilbert$ with $P_+$ the projection
$\Hilbert_\ext \to \Hilbert$, and let $U: \Hilbert_\ext \to
\int^\oplus \Hilbert_q \, dq$ be a naturalization of $\pov_\ext$.  For
every $q\in \conf$, pick an orthonormal basis $\basis_q =
\{|q,i\rangle\}$ of $\Hilbert_q$, with measurable dependence on
$q$. When each set $\basis_q$ is thought of as equipped with the
counting measure, then from $dq$ we obtain a measure on $\basis =
\bigcup_q \basis_q$, and $\int^\oplus \Hilbert_q\, dq$ is naturally
identified with $L^2(\basis,\CCC)$.  Since $H$ is a Hilbert--Schmidt
operator, so is $H_\ext= IHP_+$, which thus possesses a kernel
function $K \in L^2(\basis\times\basis, \CCC)$ such that for all
$\Phi\in\Hilbert_\ext$
\[
   U H_\ext \Phi(q,i) = \int\limits_{\conf} dq' \sum_{i' \in \basis_{q'}}
K(q,i,
   q',i') \, U\Phi(q',i').
\]
Since
\[
\sp{\Psi}{\pov(B) H \pov(C) |\Psi} = \sp{\Psi}{\pov_\ext(B) H_\ext
\pov_\ext(C) |\Psi}\,,
\]
  we have, for the same reasons as in the proof of Corollary
\ref{sta:HSQC}, that
\begin{equation}\label{HSpovint}
   \sp{\Psi}{\pov(B) H \pov(C) |\Psi} = \int\limits_{B \times C} dq \,
   dq' \sum_{i\in \basis_q} \sum_{i' \in \basis_{q'}}  U\Psi^*(q,i) \,
   K(q,i,q',i') \, U\Psi(q',i') \,.
\end{equation}
For $A \subseteq \conf \times \conf$, set
\[
   \mu(A) = \int\limits_{A} dq\, dq' \sum_{i \in \basis_q} \sum_{i' \in
   \basis_{q'}} U\Psi^*(q,i) \, K(q,i,q',i') \, U\Psi(q',i').
\]
Since $U\Psi^*(q,i) \, K(q,i,q',i') \, U\Psi(q',i')$ is absolutely
summable and integrable over $q,i,q'$, and $i'$, $\mu(A)$ is finite,
and thus a complex measure. \eqref{HSpovint} entails that
\eqref{muPsiPH} is satisfied. Thus Theorem \ref{sta:finiterates}
applies.
\end{proof}

We now provide the most general version of our statement about jump 
rates
for  Hamiltonians with kernel measures. Let $\bdd(\Hilbert_{q'},
\Hilbert_{q})$ denote the space of bounded linear operators 
$\Hilbert_{q'}
\to \Hilbert_{q}$ with the operator norm
\[
   |O| = \sup\limits_{\Phi \in\Hilbert_{q'},\Phi \neq 0}
    \frac{\|O\Phi\|}{\|\Phi\|}.
\]
For the norm of $\Psi(q)$ in $\Hilbert_q$,
$\scalar{\Psi(q)}{\Psi(q)}_q^{1/2}$, we also write $|\Psi(q)|$.

\begin{corollary}\label{sta:povkernel}
   Let $\Hilbert$ be a Hilbert space, $\Psi \in \Hilbert$ with
   $\|\Psi\| =1$, $H$ a self-adjoint operator on $\Hilbert$, $\conf$ a
   standard Borel space, and $\pov$ a POVM on $\conf$ acting on
   $\Hilbert$. Let $\pov_\ext$ be the Naimark extension PVM of $\pov$
   acting on $\Hilbert_\ext \supseteq \Hilbert$, and $U: \Hilbert_\ext
   \to \int^\oplus\Hilbert_q \, dq$ the naturalization of
   $\pov_\ext$. Suppose that $H$ has a kernel $K(dq \times dq')$ for
   $\Psi$ in the position representation defined by $\pov$; i.e.,
   suppose that $K(dq \times dq')$ is the product of a $\sigma$-finite
   nonnegative measure on $\conf \times \conf$ and a measurable
   cross-section of the field $\bdd(\Hilbert_{q'},\Hilbert_{q})$ over
   $\conf \times \conf$, that $\Psi$ satisfies \eqref{KHPsiformdomain},
   and that some everywhere-defined version $\Psi(q)$ of the
   almost-everywhere-defined cross-section $U\Psi \in \int^\oplus
   \Hilbert_q \, dq$ satisfies \eqref{KHPsifinite} and
   \eqref{KHPsiformkernel} (where the integrand on the right hand side
   of \eqref{KHPsiformkernel} is understood as involving the inner
   product of $\Hilbert_q$).  Then, by virtue of Theorem
   \ref{sta:finiterates}, the jump rates given by \eqref{tranrates} are
   well-defined and finite $\measure$-almost everywhere, and they are
   equivariant if $\Psi \in \mathrm{domain}(H)$.
\end{corollary}

The proof of Corollary \ref{sta:Ltwokernel} applies here without
changes if one understands $\Psi^*(q) \, K(dq \times dq') \, \Psi(q)$
as meaning $\scalar{\Psi(q)}{K(dq \times dq') \, \Psi(q')}_q$.
Corollary \ref{sta:povkernel} defines a set  of good $\Psi$'s, for
which the jump rates are finite,  for the examples of Sections
\ref{sec:Bell}, \ref{sec:expovm1}, and \ref{sec:expovm2}.

\subsection{Global Existence Question}

The rates $\sigma_t$ and velocities $v_t$, together with $\measure_t$,
define the process $Q_t$ associated with $H,\pov$, and $\Psi$, which
can be constructed along the lines of Section
\ref{sec:revjump}. However, the rigorous existence of this process,
like the global existence of solutions for an ordinary differential
equation, is no trivial matter. In order to establish the global
existence of the process (see \cite{crex1} for an example), a variety
of aspects must be controlled, including the following: (i)~One has to
show that for a sufficiently large set of initial state vectors, the
relevant conditions for finiteness of the jump rates, see Sections
\ref{sec:thm} and \ref{sec:jumps}, are satisfied at all
times. (ii)~One has to show that there is probability zero that
infinitely many jumps accumulate in finite time. (iii)~One has to show
that there is probability zero that the process runs into a
configuration where $\sigma$ is ill defined (e.g., where the
denominator of \eqref{mini1} vanishes, if that equation is
appropriate).

\subsection{Extensions of Bi-Measures}\label{sec:tensorpovm}

We have pointed out in the next-to-last paragraph of
Section~\ref{sec:thm} that a complex bi-measure need not possess an
extension to a complex measure on the product space, a fact relevant
to the conditions for finite rates.  In this section we show, see
Theorem~\ref{sta:bimeasure} below, that nonnegative real bi-measures
always possess such an extension.

A useful corollary of Theorem~\ref{sta:bimeasure}, see
Corollary~\ref{sta:tensorpovm} below, asserts that one can form the
tensor product of any two POVMs. This is a special case of the more
general statement, see Corollary~\ref{sta:prodpovm} below, asserting
that one can form the product of any two POVMs that commute with each
other; this statement can be regarded as the generalization from PVMs
to POVMs of the fact that two commuting observables can be measured
simultaneously; it is also related to the discussion in the last
paragraph of Section \ref{sec:thm}.

Though we could not find the explicit statement of
Corollary~\ref{sta:prodpovm} in this form in the literature, it does
follow from a part of a proof given by Halmos \cite[p.~72]{HalHil}.
Below, however, we give a somewhat different proof, using
Theorem~\ref{sta:bimeasure} instead of the lemma of von Neumann
\cite[p.~167]{vonNeumann} that Halmos uses.  It is also presumably
possible to derive Corollary~\ref{sta:prodpovm} from Lemma 2.1 or
Theorem 2.2 of \cite{Davies}.

\begin{theorem}\label{sta:bimeasure}
   Let $\conf_1$ and $\conf_2$ be standard Borel spaces with
   $\sigma$-algebras $\salg_1$ and $\salg_2$, and let $\nu( \,\cdot\,,
   \,\cdot\,)$ be a finite nonnegative bi-measure, i.e., a mapping $\nu:
   \salg_1 \times \salg_2 \to [0,a]$, $a > 0$, that is a measure
   in each variable when the other variable is a fixed set. Then $\nu$
   can be extended to a measure $\mu$ on $\conf_1 \times \conf_2$:
   there exists a unique finite nonnegative measure $\mu: \salg_1
   \otimes \salg_2 \to [0,a]$ such that for all $B_1 \in \salg_1$ and
   $B_2 \in \salg_2$,
   \begin{equation}\label{bimeasure}
     \mu(B_1 \times B_2) = \nu(B_1, B_2).
   \end{equation}
\end{theorem}

\begin{proof}
Suppose first that $\conf_1$ is finite or countably infinite. Then
every set $A \in \salg_1 \otimes \salg_2$ is an (at most) countable
union of product sets,
\[
   A = \bigcup_{q_1 \in \conf_1} \{ q_1 \} \times B_{q_1},
\]
where every $B_{q_1} \in \salg_2$. Therefore, the unique way of
extending $\nu$ is by setting
\begin{equation}\label{bimcountQ}
   \mu(A) := \sum_{q_1 \in \conf_1} \nu(\{q_1\}, B_{q_1}).
\end{equation}
One easily checks that \eqref{bimcountQ} indeed defines a finite measure
satisfying \eqref{bimeasure},
noting first that the sum is always finite because $\sum \nu(\{ q_1\},
B_{q_1}) \leq \sum \nu(\{ q_1\}, \conf_2) = \nu(\conf_1,\conf_2)$.
The same argument can of course be applied if $\conf_2$ is finite or 
countably
infinite.

Suppose now that neither $\conf_1$ nor $\conf_2$ is finite or countable.
Every uncountable standard Borel space $\conf$ is isomorphic, as a
measurable space, to the space of binary sequences $\{0,1\}^\NNN$
(equipped with the $\sigma$-algebra generated by the family
$\mathcal{B}$ of sets that depend on only finitely many terms of the
sequence), i.e., there exists a bijection $\varphi: \conf \to
\{0,1\}^\NNN$ that is measurable in both directions, see
\cite[p.~138]{Mackey} and \cite[p.~358]{Kuratowski}. We may thus assume,
without loss of generality, that $\conf_1=\{0,1\}^{\{-1,-2,-3,\dots\}}$ 
and
$\conf_2=\{0,1\}^{\{0,1,2,\dots\}}$, with $\mathcal{B}_i$ defined
accordingly. $\conf_1 \times \conf_2$ can then be canonically identified
with $\{0,1\}^\ZZZ$.

{}From the restriction of $\nu$ to sets
$B_1\in\mathcal{B}_1$ and $B_2\in\mathcal{B}_2$, one easily obtains a
consistent family of finite-dimensional distributions, and hence, by the
Kolmogorov extension theorem, e.g. \cite[p.~24]{Breiman}, a unique 
measure
$\mu$ on  $\conf_1 \times \conf_2$ obeying \eqref{bimeasure} for all 
$B_1\in\mathcal{B}_1$ and $B_2\in\mathcal{B}_2$.

It remains to establish \eqref{bimeasure} for all $B_1\in\salg_1$ and
$B_2\in\salg_2$. First fix $B_2$ in $\mathcal{B}_2$. Then $\mu(\,\cdot\,
\times B_2)$ and $\nu(\,\cdot\,, B_2)$ are measures on $\salg_1$ that 
agree
on $\mathcal{B}_1$. Hence they agree on $\salg_1$. Thus, fixing $B_1$ in
$\salg_1$, we have that $\mu(B_1\times \,\cdot\,)$ and $\nu(B_1,
\,\cdot\,)$ are measures on $\salg_2$ that agree on $\mathcal{B}_2$, and
hence on all of $\salg_2$, completing the proof.
\end{proof}

In the following, we will again write $B_1 \subseteq \conf_1$ instead
of $B_1 \in \salg_1$.

\begin{corollary}\label{sta:prodpovm}
   Let $\Hilbert$ be a Hilbert space, $\conf_1$ and $\conf_2$ standard
   Borel spaces, and $\pov_1$ and $\pov_2$ POVMs on $\conf_1$ and
   $\conf_2$ respectively, acting on $\Hilbert$. If
   $[\pov_1(B_1),\pov_2(B_2)]=0$ for all $B_1 \subseteq \conf_1$ and
   $B_2 \subseteq \conf_2$, then there exists a unique POVM $\pov$ on
   $\conf_1 \times \conf_2$ acting on $\Hilbert$ such that for all $B_1
   \subseteq \conf_1$ and $B_2 \subseteq \conf_2$,
   \begin{equation}\label{prodpovm}
     \pov(B_1 \times B_2) = \pov_1(B_1) \pov_2(B_2).
   \end{equation}
\end{corollary}

\begin{proof}
(We largely follow \cite[p.~72]{HalHil}.) For $\Psi \in \Hilbert$ we
define a bi-measure $\nu_\Psi$ by setting $\nu_\Psi(B_1, B_2) := 
\sp{\Psi}
{\pov_1(B_1) \pov_2(B_2) | \Psi}$. $\nu_\Psi$ is obviously a complex
bi-measure, and it takes values only in the nonnegative reals because
$\pov_1(B_1) \pov_2(B_2)$ is a positive operator (since the two
positive operators $\pov_1(B_1)$ and $\pov_2(B_2)$ can be
simultaneously diagonalized). The values of $\nu_\Psi$ are bounded by
$\|\Psi\|^2$. By Lemma \ref{sta:bimeasure}, $\nu_\Psi$ can be extended
to a measure $\mu_\Psi$ on $\conf_1 \times \conf_2$.

We now define complex measures $\mu_{\Phi,\Psi}$ on $\conf_1 \times
\conf_2$ by ``polarization'': for every $A \subseteq \conf_1 \times
\conf_2$ and for every pair of vectors $\Phi,\Psi$ we write
\begin{equation}
   \mu_{\Phi,\Psi}(A) := \mu_{\frac 12 \Phi + \frac 12 \Psi}(A) -
   \mu_{\frac 12 \Phi - \frac 12 \Psi}(A) + i\mu_{\frac 12 \Phi - \frac
   i2 \Psi}(A) -i\mu_{\frac 12 \Phi + \frac i2 \Psi}(A).
\end{equation}
We assert that $\mu_{\Phi,\Psi}(A)$ is, for each fixed set $A$, a
symmetric bilinear functional. This assertion is proved by noting that
(i) it is true if $A= B_1 \times B_2$, and (ii) the class of all sets
for which it is true is closed under the formation of complements and
countable unions. To see (ii), note that $\mu_{\Phi,\Psi}(A^c) =
\mu_{\Phi,\Psi}(\conf_1 \times \conf_2) - \mu_{\Phi,\Psi}(A)$ and
$\mu_{\Phi,\Psi}(\bigcup^\infty_{k=1} A_k) = \lim_{n \to \infty}
\sum_{k=1}^n \mu_{\Phi,\Psi}(A_k)$.

Since $\mu_{\Psi,\Psi}(A) = \mu_\Psi(A) \leq \|\Psi\|^2$ for every
$A\subseteq \conf_1 \times \conf_2$, the bilinear functional
$\mu_{\Phi,\Psi}(A)$ is bounded and has, in fact, a norm $\leq
1$. Therefore, there is a bounded operator $\pov(A)$ such that
$\mu_{\Phi,\Psi}(A) = \sp{\Phi} {\pov(A)| \Psi}$. $\pov(A)$ is
positive since $\mu_{\Psi,\Psi}(A) \geq 0$ for every
$\Psi$. $\pov(\,\cdot\,)$ is countably additive in the weak operator
topology because $\mu_{\Phi,\Psi}(\,\cdot\,)$ is countably additive.
$\pov(\,\cdot\,)$ satisfies \eqref{prodpovm}, and thus $\pov(\conf_1
\times \conf_2) = I$.
\end{proof}

Note that $\pov_1$ need not be a commuting POVM, i.e., possibly 
$[\pov_1(B_1),
\pov_1(C_1)] \neq 0$, and correspondingly for $\pov_2$.

An immediate consequence of Corollary \ref{sta:prodpovm}, which we use
in several places of \cite{crea2B}, is

\begin{corollary}\label{sta:tensorpovm}
   Let $\Hilbert_1$ and $\Hilbert_2$ be Hilbert spaces, $\conf_1$ and
   $\conf_2$ standard Borel spaces, and $\pov_1$ and $\pov_2$ POVMs on
   $\conf_1$ and $\conf_2$ respectively, acting on $\Hilbert_1$ and
   $\Hilbert_2$ respectively. Then there exists a unique POVM $\pov$ on
   $\conf_1 \times \conf_2$ acting on $\Hilbert_1 \otimes \Hilbert_2$
   such that for all $B_1 \subseteq \conf_1$ and $B_2 \subseteq
   \conf_2$,
   \begin{equation}\label{tensorpovm}
     \pov(B_1 \times B_2) = \pov_1(B_1) \otimes \pov_2(B_2).
   \end{equation}
\end{corollary}

\section{Minimality}
\label{sec:mini4}

In this section we explain in what sense the minimal jump rates
\eqref{tranrates}---or \eqref{mini1} or \eqref{disrates1}---are
minimal.  In so doing, we will also explain the significance of the
quantity $\current$ defined in \eqref{Jdef}, and clarify the meaning
of the steps taken in Sections \ref{sec:mini1} and \ref{sec:mini2} to
arrive at the jump rate formulas.

Given a Markov process $Q_t$ on $\conf$, we define the \emph{net
probability current} $j_t$ at time $t$ between sets $B$ and $B'$ by
\begin{eqnarray}\label{jdefcont}
     j_t(B,B') = \lim_{\Delta t \searrow 0} \frac{1}{\Delta t}
\hspace{-3ex}
   &&
     \Big[ \prob\big\{Q_{t}\in B',Q_{t+\Delta t}\in B  \big\} -
   \\\nonumber
   &&
   - \prob \big\{ Q_{t}\in B, Q_{t+\Delta t} \in B' \big\} \Big]\,.
\end{eqnarray}
This is the amount of probability that flows, per unit time, from $B'$
to $B$ minus the amount from $B$ to $B'$.  For a pure jump process, we
have that
\begin{equation}\label{jrate}
   j_t(B,B') = \int\limits_{q'\in B'} \sigma_t(B|q')\, \rho_t(dq') -
   \int\limits_{q\in B} \sigma_t(B'|q)\, \rho_t(dq)\,,
\end{equation}
so that
\begin{equation}
j_t(B,B') = j_{\sigma,\rho}(B \times B')
\end{equation}
where $j_{\sigma,\rho}$ is the signed measure, on $\conf \times
\conf$, given by the integrand of \eqref{continuity3},
\begin{equation}\label{jsigma}
   j_{\sigma,\rho} (dq \times dq') = \sigma(dq|q') \, \rho(dq') -
   \sigma(dq'|q) \, \rho(dq)\,.
\end{equation}
For minimal jump rates $\sigma$, defined by \eqref{tranrates} or
\eqref{mini1} or \eqref{disrates1} (and with the probabilities $\rho$
given by \eqref{mis}, $\rho = \measure$), this agrees with \eqref{Jdef},
as was noted earlier,
\begin{equation}\label{jJ}
   j_{\sigma,\rho} = \current_{\Psi,H,\pov} \,,
\end{equation}
where we have made explicit the fact that $\current$ is defined in
terms of the quantum entities $\Psi, H$, and $\pov$. Note that both
$\current$ and the net current $j$ are anti-symmetric, $\current^\tr =
-\current$ and $j^\tr = -j$, the latter by construction and the former
because $H$ is Hermitian. (Here $\tr$ indicates the action on measures
of the transposition $(q,q') \mapsto (q',q)$ on $\conf \times \conf$.)
The property \eqref{jJ} is stronger than the equivariance of the rates
$\sigma$, $\generator_\sigma \measure_t = d\measure_t / dt$: Since, by
\eqref{continuity3},
\begin{equation}
   (\generator_\sigma \rho) (dq) = j_{\sigma,\rho} (dq \times \conf),
\end{equation}
and, by \eqref{Jdef},
\begin{equation}
   \frac{d\measure}{dt}(dq) = \current(dq \times \conf),
\end{equation}
the equivariance of the jump rates $\sigma$ amounts to the condition
that the marginals of both sides of \eqref{jJ} agree,
\begin{equation}
   j_{\sigma,\rho} (dq \times \conf) = \current (dq \times \conf)\,.
\end{equation}
In other words, what is special about processes with rates satisfying
\eqref{jJ} is that not only the single-time \emph{distribution} but
also the \emph{current} is given by a standard quantum theoretical
expression in terms of $H, \Psi$, and $\pov$. That is why we call
\eqref{jJ} the \emph{standard-current property}---defining
\emph{standard-current rates} and \emph{standard-current processes}.

Though the standard-current property is stronger than equivariance, it
alone does not determine the jump rates, as already remarked in
\cite{BD,Roy}. This can perhaps be best appreciated as follows: Note
that \eqref{jsigma} expresses $j_{\sigma,\rho}$ as twice the
anti-symmetric part of the (nonnegative) measure
\begin{equation}
  C(dq \times dq') = \sigma(dq|q') \, \rho(dq')
\end{equation}
on $\conf \times \conf$ whose right marginal $C(\conf \times dq')$ is
absolutely continuous with respect to $\rho$. Conversely, from any
such measure $C$ the jump rates $\sigma$ can be recovered by forming
the Radon--Nikod\'ym derivative
\begin{equation}
   \sigma(dq|q') = \frac{C(dq \times dq')}{\rho(dq')}\,.
\end{equation}
Thus, given $\rho$, specifying $\sigma$ is equivalent to specifying
such a measure $C$.

In terms of $C$, the standard-current property becomes (with $\rho =
\measure$)
\begin{equation}\label{CJ}
   2 \, \mathrm{Anti} \, C = \current.
\end{equation}
Since (recalling that $\current = \current^+ - \current^-$ is
anti-symmetric)
\begin{equation}
   \current = 2 \, \mathrm{Anti} \, \current^+,
\end{equation}
an obvious solution to \eqref{CJ} is
\[
   C = \current^+,
\]
corresponding to the minimal jump rates. However, \eqref{jJ} fixes
only the anti-symmetric part of $C$. The general solution to
\eqref{CJ} is of the form
\begin{equation}
   C = \current^+ + S
\end{equation}
where $S(dq \times dq')$ is symmetric, since any two solutions to
\eqref{CJ} have the same anti-symmetric part, and $S \geq 0$, since $S
= C \wedge C^\tr$, because $\current^+ \wedge (\current^+)^\tr =0$.

In particular, for any standard-current rates, we have that
\begin{equation}\label{minimality}
   C \geq \current^+, \quad \text{or} \quad \sigma(dq|q') \geq
   \frac{\current^+(dq \times dq')}{\measure(dq')}.
\end{equation}
Thus, among all jump rates consistent with the standard-current
property, one choice, distinguished by equality in \eqref{minimality},
has the least frequent jumps, or the smallest amount of stochasticity:
the minimal rates \eqref{tranrates}.

\section{Remarks}\label{sec:remarks}

\subsection{Symmetries}\label{sec:symm}

   Quantum theories, and in particular QFTs, often have important
   symmetries. To name a few examples: space translations, rotations
   and inversion, time translations and reversal, Galilean or Lorentz
   boosts, global change of phase $\Psi \to \E^{\I\theta} \Psi$, and
   gauge transformations.

   This gives rise to the question whether the process $Q_t$ of the
   corresponding Bell-type QFT respects these symmetries as well.
   Except for Lorentz invariance, which is difficult in that Lorentz
   boosts fail to map equal-time configurations into equal-time
   configurations, the answer is yes; a discussion is given in
   \cite[Sec.~6.1]{crea2B}. An essential ingredient of this result is
   the manifest fact that the minimal jump rates \eqref{tranrates}
   inherit the symmetries of the Hamiltonian (under which the POVM
   transforms covariantly).

\subsection{Homogeneity of the Rates}

   The minimal jump rates (\ref{tranrates}) define a homogeneous
   function of degree 0 in $\Psi$, i.e., $\sigma^{\lambda\Psi} =
   \sigma^{\Psi}$ for every $\lambda\in \CCC \setminus \{0\}$. This
   property is noteworthy since it forms the essential mathematical
   basis for a number of desirable properties of theories using such
   jump rates (such as that of \cite{crea1}): (i)~that (when $\pov$ is
   a product PVM) unentangled and decoupled subsystems behave
   independently and follow the same laws as the entire system,
   (ii)~that ``collapsed-away,'' i.e., sufficiently distant, parts of
   the wave function do not influence the future behaviour of the
   configuration $Q_t$, (iii)~invariance under a global change of phase
   $\Psi \to \E^{\I\theta}\Psi$, (iv)~invariance under the replacement
   $\Psi \to \E^{-\I Et/\hbar} \Psi$ for some constant $E$, which
   corresponds to adding $E$ to the total Hamiltonian, (v)~invariance
   under relabeling of the particles (which may cause a replacement
   $\Psi \to -\Psi$ due to the Pauli principle).

\subsection{$H+E$}

   Adding a constant $E$ to the interaction Hamiltonian will not
   change the jump rates \eqref{tranrates} provided $\pov$ is a PVM.
   This is because $\sp{\Psi}{\pov(B) E \pov(C) |\Psi} = E \,
   \sp{\Psi}{\pov(B \cap C) |\Psi}$ has vanishing imaginary part.  For
   a POVM, however, this need not be true.

\subsection{Nondegenerate Eigenstates}

   As mentioned earlier, after
   \eqref{mini1}, it is a consequence of the minimal jump rate formula
   (\ref{tranrates}), in fact of the very minimality, that at each time
   $t$ either $\sigma(q|q')$ or $\sigma(q'|q)$ is zero. It follows that
   for a time-reversible Hamiltonian $H$ and POVM $\pov$, all jump
   rates vanish if $\Psi$ is a nondegenerate eigenstate of $H$.  This
   is because, in the simplest cases, $\sp{q}{H|q'}$ is real, and the
   coefficients $\sp{q}{\Psi}$ can also be chosen real, or, more
   generally and more to the point, because in this case the process
   must coincide with its time reverse, which implies that the current
   from $q$ to $q'$ is as large as the one from $q'$ to $q$, so that
   minimality requires both to vanish.

\subsection{Left or Right Continuity}

   {}From what we have said so far, there
   remains an ambiguity as to whether $Q_t$ at the jump times should be
   the point of departure or the destination, in other words, whether
   the realization $t\mapsto Q_t$ should be chosen to be left or to be
   right continuous. Although we think there is not much physical
   content to this question, we should point out that demanding either
   left or right continuity will destroy time-reversal invariance (cf.\
   Section \ref{sec:symm}).  A prescription that preserves
   time-reversal invariance can, however, be devised provided the
   possible jumps can be divided into two classes, $A$ and $B$, in such
   a way that the time reverse of a class-$A$ jump necessarily belongs
   to class $B$ and vice versa. Then class-$A$ jumps can be chosen left
   continuous and class-$B$ as right continuous. An example is provided
   by the model of \cite{crea1}: since at every jump the number of
   particles either increases or decreases, the jumps naturally form
   two classes (``creation'' and ``annihilation''), and the time
   reverse of a creation is an annihilation. The prescription could be
   that if a particle is created (annihilated) at time $t$, then $Q_t$
   already (still) contains the additional particle. But the opposite
   rule would be just as consistent with time-reversal symmetry, and we
   can see no compelling reason to prefer one rule over the other.

\section{Conclusions}

We have investigated the possibility of understanding QFT as a theory
about moving particles, an idea pioneered, in the realm of
nonrelativistic quantum mechanics, by de Broglie and Bohm. The models
proposed by Bell \cite{BellBeables} and ourselves \cite{crea1} turn
out to be rather universal; that is, their construction can be
transferred to a variety of situations, involving different
Hamiltonians and configuration spaces, and invoking formulas of a
canonical character.

One ingredient of the construction is the use of stochastic jumps
whose rates are determined by the quantum state vector (and the
Hamiltonian).  These rates can be specified through an explicit
formula (\ref{tranrates}) that has a status similar to the velocity
formula in Bohmian mechanics.  We have provided a version of this jump
rate formula that is more general than any previous one. Indeed, it
seems to be the most general version possible: we need assume merely
that the configuration space $\conf$ is a measurable space (the
weakest notion of ``space'' available in mathematics), that the
Hamiltonian is well-defined, and that $\conf$ and the Hilbert space
are related through a generalized position observable (a
positive-operator-valued measure, or POVM, the most general notion
available in quantum theory of how a vector in Hilbert space may
define a probability distribution). We have shown that these jump
rates are well-defined and finite if the interaction Hamiltonian
possesses a sufficiently regular kernel in the position representation
defined by the POVM.

We have also indicated that in a Bell-type QFT, the different
contributions to the Hamiltonian correspond to different contributions
to the motion of the configuration $Q_t$.  The relevant fact is
process additivity, i.e., that the generator of the Markov process
$Q_t$ is additive in the Hamiltonian.  The free process usually
consists of continuous trajectories, Bohmian or similar, an
observation already made in \cite{crea1} for the model considered
there. Exploiting process additivity, we obtain that $Q_t$ is
piecewise deterministic, the pieces being Bohm-type trajectories,
interrupted by stochastic jumps.  Given a Hamiltonian and POVM, our
prescription determines the Markov process $Q_t$. As an example, we
have described the process explicitly for a simple QFT.

The essential point of this paper is that there is a direct and
natural way---a canonical way---of devising a Bell-type version of any
QFT.

\bigskip

\noindent \textbf{Acknowledgements. }We thank Ovidiu Costin, Avraham
Soffer, and James Taylor of Rutgers University, Stefan
Teufel of Technische Universit\"at M\"unchen, and Gianni Cassinelli
and Alessandro Toigo of Universit\`a di Genova for helpful
discussions.  R.T.\ gratefully acknowledges support by the German
National Science Foundation (DFG).  N.Z.\ gratefully acknowledges
support by INFN and DFG.  Finally, we appreciate the hospitality that
some of us have enjoyed, on more than one occasion, at the
Mathematisches Institut of Ludwig-Maximilians-Universit\"at M\"unchen,
at the Dipartimento di Fisica of Universit\`a di Genova, and at the
Mathematics Department of Rutgers University.

\end{document}